\documentclass{aa}

\usepackage{graphicx}
\usepackage{txfonts}
\usepackage{hyperref}
\hypersetup{
 colorlinks=true,
 linkcolor=blue,
 citecolor=blue
}

\newcommand{\cmt}{cm$^{-3}$}
\newcommand{\cmd}{cm$^{-2}$}

\newcommand{\nh}{$n_\text{H}$}

\begin{document} 

\title{Influence of galactic arm scale dynamics on the molecular composition of the cold and dense ISM}

\subtitle{I. Observed abundance gradients in dense clouds}

\author{M. Ruaud \inst{1} \inst{2} \and
             V. Wakelam \inst{2} \and
             P. Gratier \inst{2}\and
             I. A. Bonnell \inst{3}
            }

\institute{NASA Ames Research Center, Moffett Field, CA, USA
              \and
              Laboratoire d'astrophysique de Bordeaux, Univ. Bordeaux, CNRS, B18N, all\'ee Geoffroy Saint-Hilaire, 33615 Pessac, France
              \and
              Scottish Universities Physics Alliance (SUPA), School of Physics and Astronomy, University of St. Andrews, North Haugh, \\ St Andrews, Fife KY16 9SS, UK
             }

\date{Received September 15, 1996; accepted March 16, 1997}

\abstract{}
{We study the effect of large scale dynamics on the molecular composition of the dense interstellar medium during the transition between diffuse to dense clouds.}
{We followed the formation of dense clouds (on sub-parsec scales) through the dynamics of the interstellar medium at galactic scales. We used results from smoothed particle hydrodynamics (SPH) simulations from which we extracted physical parameters that are used as inputs for our full gas-grain chemical model. In these simulations, the evolution of the interstellar matter is followed for $\sim50$ Myr. The warm low-density interstellar medium gas flows into spiral arms where orbit crowding produces the shock formation of dense clouds, which are held together temporarily by the external pressure.}
{We show that depending on the physical history of each SPH particle, the molecular composition of the modeled dense clouds presents a high dispersion in the computed abundances even if the local physical properties are similar. We find that carbon chains are the most affected species and show that these differences are directly connected to differences in (1) the electronic fraction, (2) the C/O ratio, and (3) the local physical conditions. We argue that differences in the dynamical evolution of the gas that formed dense clouds could account for the molecular diversity observed between and within these clouds.}
{This study shows the importance of past physical conditions in establishing  the chemical composition of the dense medium.}

\keywords{Astrochemistry -- ISM: clouds -- ISM: evolution -- ISM: kinematics and dynamics -- ISM: molecules -- Galaxy: evolution}

\maketitle

\section{Introduction}
\label{sec:intro}

Observations of cold and dense interstellar clouds have revealed an extraordinary molecular richness, but also a large dispersion in the inferred abundances between and within these clouds. One of the most famous examples is given by the observation of the two low-mass cores TMC-1 and L134N for which more than 50 molecules have been identified \citep{Bergin07,Agundez13}. Abundance gradients along the TMC-1 ridge were reported almost 40 years ago and their understanding has remained an unsolved problem since then. In this source, abundance gradients were observed along the ridge at scales of few 0.1 pc \citep{Pratap97}. The southeastern end of the ridge (also known as the CP peak\footnote{CP stands for cyanopolyyne here.}) is characterized by enhanced abundances of carbon chains as compared to the northwestern end (also known as the NH$_3$ peak), while some other molecules (such as NH$_3$) are relatively well distributed along the ridge \citep{Pratap97}. For instance, HC$_7$N/NH$_3$ ratio varies by more than a factor of 30 along the ridge \citep{Olano88}. At the CP peak position the inferred abundances of complex carbon chains are typically a factor of 10 higher than what is observed in the L134N dense cloud, which is thought to have similar physical properties as TMC-1 \citep{Agundez13}. Even though L134N do not show strong abundance variations on scales comparable to TMC-1 \citep{Dickens00}, small scale abundance gradients have also been revealed (on scales of few 0.01 pc) in the dense interior of this source, suggesting the presence of a pre-stellar core \citep{Pagani03,Pagani04,Pagani05}. 

In most cases, understanding these abundance gradients is a difficult task because, first, the chemistry is out of equilibrium in most of these objects and, second, molecular observations provide us only a snapshot of their state. Although considerable theoretical work has been conducted to understand the chemistry at work in such regions \citep[e.g.,][]{Watson76,Herbst73,Prasad80,Graedel82,Leung84}, most of the time dependent chemical modeling of dense clouds relies on simple approximations that reflect these difficulties. Based on the lack of information on their dynamical state, dense clouds are often considered very simple objects in which the physical conditions are homogeneous and fixed. While such models have been shown to lead to relatively good agreement with observations in dense clouds \citep{Wakelam10,Agundez13}, they also have been highly criticized. Most of these studies neglect the complex morphology of the source, often using a zero-dimensional approximation (single point calculation with homogenous and fixed physical conditions). Such approximations can, of course, only be made up to a certain degree of spatial resolution (i.e., as revealed by the careful analysis of L134N by \citet{Pagani03,Pagani04,Pagani05}). 
Another critical point of these models is that they rely on a somewhat artificial choice of initial conditions \citep{Gerin03,Liszt09}. A good example is given by the arbitrarily low abundance of metals commonly used in these models. \citet{Graedel82} early recognized the need for lowering the electron abundance to better reproduce observations in some dense clouds. This has been achieved by decreasing the abundance of metals by a factor of $\sim 100$. This set of elemental abundances, known as low metal elemental abundances, are commonly used to model dense clouds and are aimed at taking into account the effect of the depletion during the transition between the diffuse and dense medium. However, the strength of the depletion used in this set of elemental abundances is somewhat arbitrary and does not rely on observational or theoretical constraints. One other important point discussed by \citet{Liszt09} is the absence of inherited molecules from the diffuse medium in the starting conditions of these models, for which all the elements are considered to be in their atomic form except for hydrogen, which is usually considered to be in the form of H$_2$. Indeed, while diffuse clouds have long been thought to be devoid of molecules (which could have supported the assumptions described above), it is now clear that a wide range of molecules are present in the diffuse medium \citep[e.g.,][]{Snow06,Gerin16}. This is all the more important given that the use of these initial conditions have raised the controversial idea of an early-time chemistry in dense clouds in which some objects may be seen in an early evolutionary stage. This so-called early-time chemistry is often invoked to explain the unusual large abundances of carbon chains in TMC-1 \citep{Herbst89,vanDishoeck93}.

These assumptions have also been questioned by recent hydrodynamical simulations, which have brought new insights into the dynamic governing the formation of dense clouds \citep[see, e.g.,][]{Hennebelle12,Dobbs14,Klessen16}. These studies have profoundly changed our minds about the very definition of what a dense cloud could be. Indeed all these studies show that rather than discrete objects, which could have been the case in the past, molecular clouds are constantly evolving objects that show a strong coupling with their environment \citep{Klessen16}. Although much effort has been made to couple the chemistry and dynamics, most hydrodynamical modeling of the formation of molecular clouds includes a very simple chemistry consisting of few elements and reactions \citep[e.g.,][]{Bergin04,Glover10,Clark12}. While this can be easily understood considering the computational requirements of such simulations, the impact of the dynamical evolution of the medium on the molecular content of dense clouds is poorly constrained and focuses on simple polyatomic molecules and ions \citep{Glover10,Clark12}.

Some studies have proposed overcoming this limitation by computing the chemistry as a post-treatment of hydrodynamical simulations \citep{Hassel10,Hincelin13}. Such hybrid models allow the use of complex chemical models, which include several hundreds of chemical species coupled, through thousands of reactions, with outputs from complex high-resolution hydrodynamical simulations.

In this study, we adopt such technique to follow the evolution of the chemical composition during the formation process of dense clouds from the diffuse medium. To do so, we follow the formation of dense clouds (on sub-parsec scales) through the dynamics of the interstellar medium at galactic scales using results of SPH smoothed particle hydrodynamics (SPH) simulations from \citet{Bonnell13}. We then post-process these results with our full gas-grain chemical model Nautilus \citep{Ruaud16} to get the evolution of the chemical composition. The main difference from the other more classical approaches to studying the chemical composition of dense clouds is that we do not make any assumptions about the initial composition of the dense cloud and let the chemical model evolve continuously from the diffuse to the dense interstellar medium. 

The paper is organized as follows. In Section \ref{sec:method} we give an overview of the hydrodynamical simulations that we use, and present our cloud selection and our chemical model Nautilus. In Section \ref{sec:results} we present the results obtained for the two dense clouds that we selected. In Section \ref{sec:discussion} we discuss these results and propose an analogy with what is observed and Section \ref{sec:conclusion} contains a summary of our work.

\section{Method}
\label{sec:method}

\subsection{Hydrodynamical simulations overview}

The numerical simulations of \citet{Bonnell13} uses a three-dimensional SPH method to follow the gas dynamics in a galactic potential including spiral arms. A top-down approach is used to probe successive smaller scales from the scale of a spiral galaxy to the scale of dense cloud formation. In an initial simulation, the full galactic disk dynamics is followed over a period of $\sim 360$ Myr. This simulation uses $2.5 \times 10^7$ SPH particles to model $10^9$ M$_\odot$ of gas in an annulus 5 to 10 kpc. A high-resolution simulation is then performed for the last 54 Myr of the full disk evolution. This simulation was performed by selecting a $250  \times 250$ pc$^2$ region at the end of the full disk simulation and by splitting original SPH particles into 256 lower mass particles. For this simulation, the selected region was traced backward through 54 Myr to produce the initial conditions for the high-resolution re-simulation. The selected region contains $1.09\times10^7$ SPH particles with an individual particle mass of 0.156 M$_\odot$. In this paper we focus on high-resolution re-simulations and define $t=0$ Myr as the starting time of the simulation such that this time corresponds to $t=310$ Myr of the full disk simulation \citep[see Fig. 1 and 2 of][]{Bonnell13}. In the selected high-resolution re-simulation region, gas from the spiral arm is cold ($T_\text{gas} \lesssim 100$ K) and dense ($n_\text{H} \gtrsim 100$ \cmt), whilst gas entering comprises warm gas ($T_\text{gas} \gtrsim 8000$ K) and some cooler and denser gas from previous spiral arm encounters. Dense clouds are formed by converging flows induced by shocks when the matter enters the spiral arm potential. At each shock, the gas is compressed and cools while maintaining pressure equilibrium with the surrounding warm gas \citep{Bonnell13}. Self-gravity is not included in these simulations such that if the external pressure decreases, the cool gas can re-expand and warm up.

\subsection{Cloud selection and physical history}

We performed cloud selection on the results obtained in the high-resolution re-simulation. This selection was carried out by first selecting a high density SPH particle in the dense cloud regime (typically with \nh~$>10^5$ \cmt). The cloud was subsequently defined by looking at all the particles passing in a radius $\le 0.5$ pc of the position of this first selected particle at its maximum density. The selected clouds are composed of few $10^2$ SPH particles and have a mass of few $10$ M$_\odot$. The physical history (e.g., temperature and density) of all the SPH particles that compose each cloud was then extracted and used as input of our full gas-grain chemical model.

The visual extinction $A_\text{V}$ has been estimated using the standard conversion from the total hydrogen column density, $N_\text{H} = N(\text{H}) + 2N(\text{H}_2)$ \citep{Bohlin78,Rachford09}, written as\begin{equation}
A_\text{V} = \frac{R_\text{V}}{C_\text{D}} \Bigg( \frac{N_\text{H}}{1~\text{cm}^{-2}} \Bigg)
,\end{equation}
where $R_\text{V} = 3.1$ and $C_\text{D} = 5.8 \times 10^{21}$ \cmd~mag$^{-1}$.
In a standard estimation of $A_\text{V}$, the knowledge of $N_\text{H}$ requires a selected a line of sight on which the integration over \nh~ is made. In our case, $A_\text{V}$ is estimated for each SPH particle by assuming $N_\text{H} = h n_\text{H}$, where $h$ is the smoothing length of the considered SPH particle. 

The dust temperature is computed by the chemical model following the approach presented in \citet{Hollenbach91}. Dust absorption $Q_\text{abs}$ was taken from \citet{Draine84} for 0.1$\mu$m silicate grains. The considered interstellar radiation field is composed by (1) the UV background \citep[taken from][]{Mathis83}, (2) the infrared emission from dust, and (3) the cosmic microwave background. The infrared emission spectrum from dust was taken from \citet{Draine07} and corresponds to the model with $R_V=3.1$, $q_\text{PAH}=4.6\%$ and $\chi = 1$. Using this approach the dust temperature ranges between $T_\text{dust}\approx 16$K in unshielded environment and $T_\text{dust}\approx 10$K when $A_V \approx$ 10.

\begin{figure*}
  \centering
  \includegraphics[width=15cm]{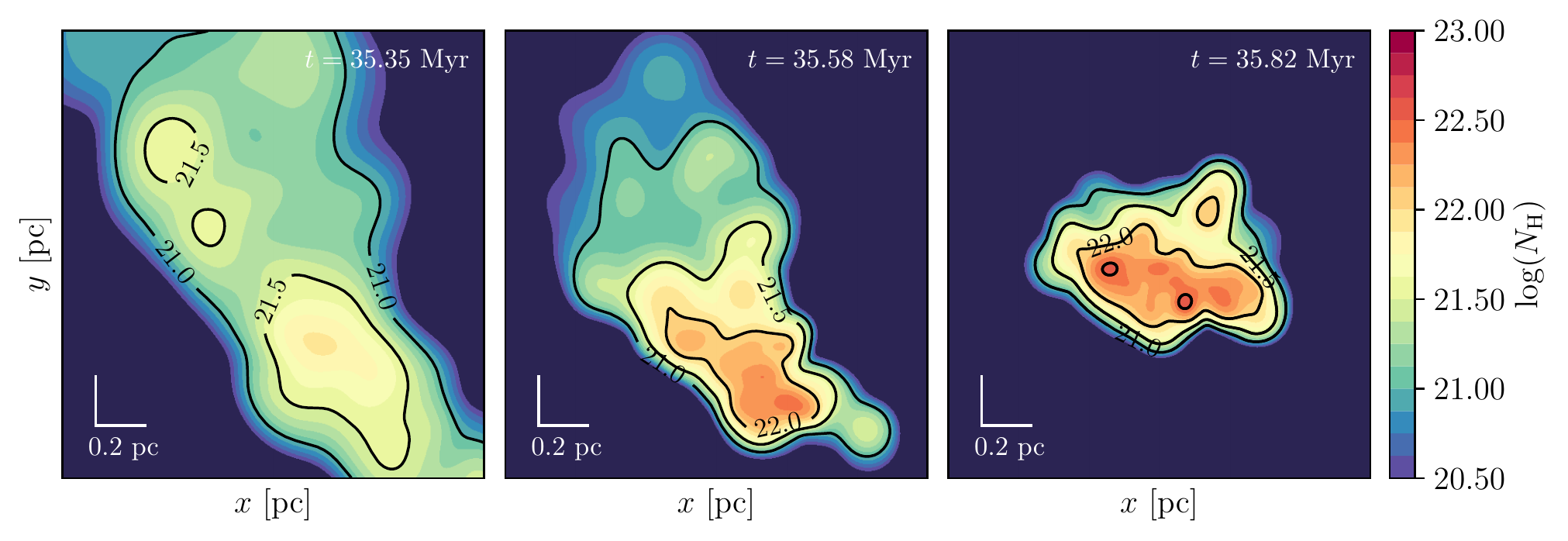}
    \caption{Evolution of the total column density map of the cloud (denoted as cloud A in the following) at $t=35.35$ Myr, $t=35.58$ Myr, and $t=38.82$ Myr, respectively. This cloud is formed by $\approx 250$ SPH particles and dissipates after $t\approx 36$ Myr.}
    \label{fig:clump_11_evol}
\end{figure*}


\begin{figure}
  \centering
  \includegraphics[width=9.2cm]{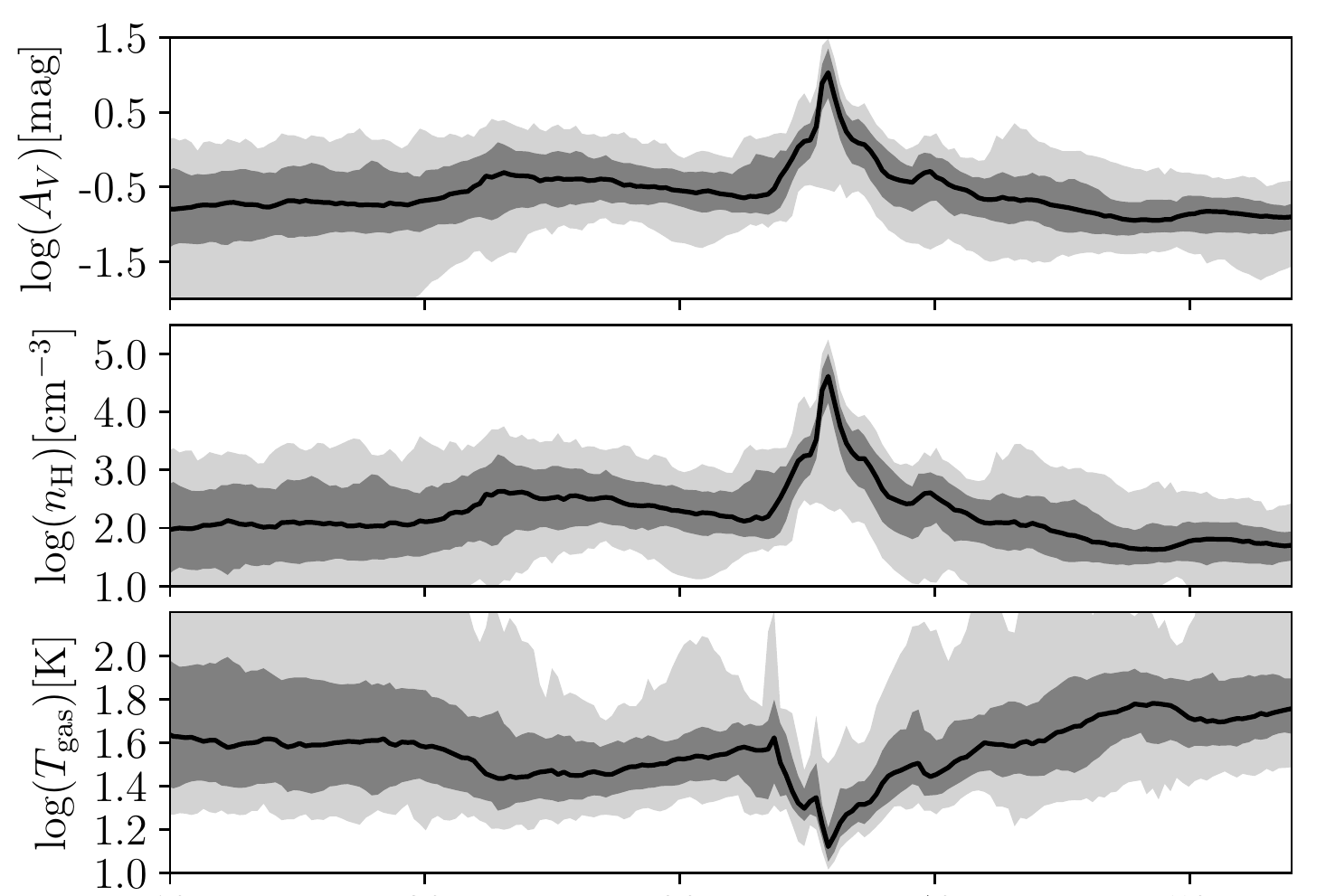}
  \includegraphics[width=9.2cm]{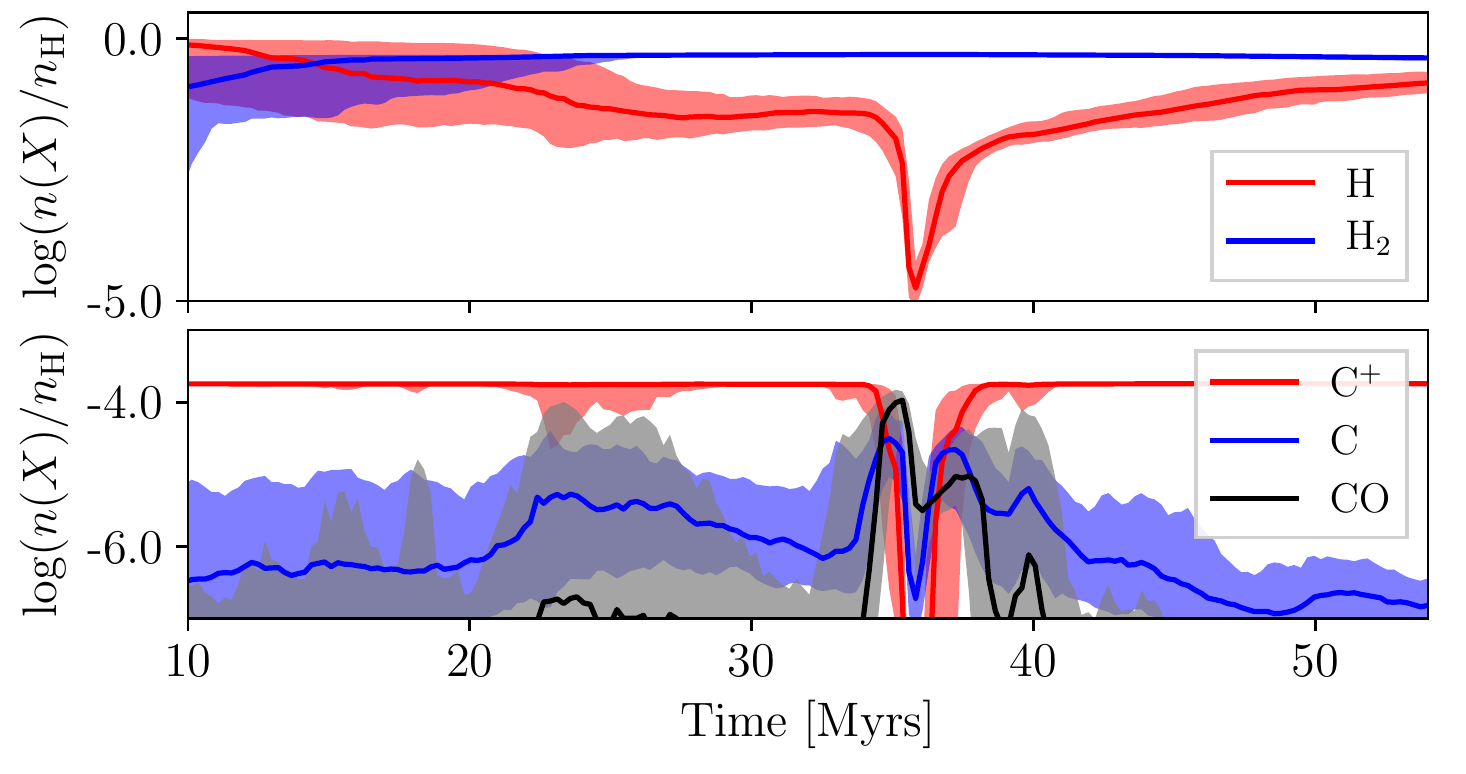}
    \caption{Statistics associated with the evolution of the physical parameters of all SPH particles that compose cloud A. For the three first plots, the black line shows the evolution of the median of the physical parameters. The dark gray filled area shows the 16th and 84th percentiles and the light gray filled area indicates the 2nd and 98th percentiles of the distribution. For the last two plots, the lines show the evolution of the median of the abundances while the filled area shows the 16th and 84th percentiles of the distribution.}
    \label{fig:clump_11_stat}
\end{figure}

Figures \ref{fig:clump_11_evol} and \ref{fig:clump_11_stat} illustrate this selection for one cloud (denoted as cloud A in the following) composed of $\sim 250$ SPH particles. Figure \ref{fig:clump_11_evol} presents the evolution of the total column density map integrated along the $z$-axis (i.e., its morphology) of the cloud at $t=35.35$ Myr, $t=35.58$ Myr, and $t=35.82$ Myr, respectively. The last map (i.e., at $t=38.82$ Myr) corresponds to the peak density of the cloud on which the selection was made. Figure \ref{fig:clump_11_stat} presents the statistics associated with the evolution of the physical parameters extracted for all the SPH particles that compose this cloud and their evolution as a function of time. It also presents the computed evolution of H, H$_2$, and C$^+$, C, CO, which are discussed in more detail in Section \ref{sec:results}. As shown in Fig. \ref{fig:clump_11_stat}, the cloud dissipates after the peak density is reached. As previously discussed, this comes from the fact that self-gravity is not taken into account in these simulations. In the following, we only focus on all the phases upstream of the cloud dissipation and ignore the downstream phases.

\subsection{Post-processed chemistry}

Each physical history was then used as an input for our gas-grain chemical model Nautilus \citep{Ruaud16}. In this model, we computed the abundance of each species by solving a set of rate equations for gas-phase and grain-surface chemistries. The chemical network used is based on {\it kida.uva.2014} \citep{Wakelam15} and includes the chemical schemes for carbon chains proposed by \citet{Wakelam15b,Loison16,Hickson16} and the update of the sulfur chemistry network presented in \citet{Vidal17}. This model computes the abundance of $\sim 900$ chemical species ($\sim 600$ for the gas phase and $\sim 300$ for the grain surface) that are coupled through more than 12\,000 reactions. Gas-phase and grain-surface chemistries are connected via accretion and desorption. For grain chemistry the model was used in its three-phase configuration, which means that a distinction is made between the surface layers (with at most two monolayers of adsorbed molecules) and the bulk. In this model,  the surface and bulk are considered as chemically active and we assume that the bulk diffusion is driven by the diffusion of water molecules \citep[see][for a complete presentation of the model]{Ruaud16}. Finally, grains are considered to be spherical with a 0.1 $\mu$m radius and a dust-to-gas mass ratio of $0.01$. For all the chemical models presented throughout this paper, we used initial abundances relevant for diffuse cloud conditions. Initial abundances are listed in Table \ref{tab:init_ab} and correspond to abundances observed toward $\zeta$ Oph ($v_\odot = -15$ km.s$^{-1}$) listed in \citet{Jenkins09}. These initial conditions differ from the low-metal abundances typically used in pseudo-time-dependent chemical models in several aspects. Firstly, no assumptions were made about the depletion of metals and their abundances are higher by at least 2 dex as compared to \citet{Graedel82}. Secondly, the initial abundances of N, O, and C$^+$ are higher by 0.45 dex, 0.27 dex and 0.4 dex, respectively, leading to an initial C/O ratio of $\approx0.55$. The cosmic ray ionization rate was taken to be equal $\zeta_{\text{H}_2} = 1.3\times10^{-17}$ s$^{-1}$.

\begin{table}
   \begin{center}
   \begin{tabular}{lrlr}
   \hline
   \hline
   Species                      &       Value   &Species                        &       Value \\
   \hline
   H                            &       1.0             			&   Fe$^+$              &       $2.0\times10^{-7}$  \\ 
   He                           &       0.09           			 &   Na$^+$              &       $2.3\times10^{-7}$  \\  
   N                            &       $6.2\times10^{-5}$        	&   Mg$^+$              &       $2.3\times10^{-6}$  \\
   O                            &       $3.3\times10^{-4}$ 		&   P$^+$                &       $7.8\times10^{-8}$          \\           
   C$^+$                        &       $1.8\times10^{-4}$ 	&   Cl$^+$                      &       $3.4\times10^{-8}$          \\      
   S$^+$                        &       $1.5\times10^{-5}$ 	&   F                           &       $1.8\times10^{-8}$          \\           
   Si$^+$                       &       $1.8\times10^{-6}$ \\
   \hline
   \end{tabular}
   \end{center}
    \caption{Initial abundances.}
   \label{tab:init_ab}
\end{table}

\section{Results}
\label{sec:results}

In the following, we focus on two dense clouds (i.e., clouds A and B) among our selection. The selection of these two clouds has been motivated by (1) the similarity of the physical parameters in the dense cloud regime, (2) the observed difference in the physical parameters in the phases that precede their formation, and (3) the fact that they illustrate the two most extreme case of all the selected clouds. Table \ref{tab:cloud_param} summarizes the physical parameters of each cloud at their peak density.

\begin{table}
   \begin{center}
   \begin{tabular}{lrr}
   \hline
   \hline
                                                        &       Cloud A & Cloud B         \\
   \hline
   Number of SPH particles              &       237               &   287               \\ 
   Total mass (M$_\odot$)$^a$           &       37        &   45                \\ 
   Mean radius (pc)                             &       0.33              &   0.29               \\
   Velocity dispersion  (km.s$^{-1}$)   &       1.4               &   0.7               \\
   Virial parameter$^b$                 &       20.0              &   4.0               \\
   Median $T_\text{gas}$ (K)            &       12.0              &  11.0               \\
   Median $n_\text{H}$ (\cmt)           & $4\times10^4$&$7\times10^4$\\
   \hline
   \end{tabular}
   \end{center}
    \caption{Physical parameters of the two clouds considered in this study. These parameters are those obtained at the peak density of each clouds. $^a$ The total mass is obtained by multiplying the number of SPH particles that compose each cloud by the individual particle mass 0.156 M$_\odot$. $^b$ The virial parameter of each cloud has been estimated  by $\alpha=5\sigma_v^2R/GM$ \citep{Bertoldi92}.}
    \label{tab:cloud_param}
\end{table}

\subsection{General results illustrated in the example of the typical cloud A}

This section gives an overview of the results obtained for cloud A presented in Figs. \ref{fig:clump_11_evol} and \ref{fig:clump_11_stat}. As shown in Fig. \ref{fig:clump_11_stat}, in the phases that precede the formation of the dense cloud, most of the SPH particles have typical diffuse and translucent gas physical conditions \citep[following the classification of][]{Snow06}. Between $t=0$ and $t=33$ Myr, the median density is $n_\text{H} \approx 100$ \cmt, the median gas temperature is $T_\text{gas}\approx 40$ K, and the median visual extinction is $A_\text{V} \approx 0.2$ mag (as shown in Fig. \ref{fig:clump_11_stat}, it is important to note the large dispersion around these values). In this simulation, the H-H$_2$ transition occurs around $t\approx 15$ Myr. After this time, the molecular fraction, $f_{\text{H}_2} = 2n(\text{H}_2) / [n(\text{H}) + 2n(\text{H}_2)]$ , slowly increases from $f_{\text{H}_2} \approx 0.5$ to a plateau of $f_{\text{H}_2} \approx 0.9$ at $t\approx25$ Myr, where $n_\text{H}\approx 300$ \cmt~ until it reaches $f_{\text{H}_2} \approx 1$ when $n_\text{H} \gtrsim 10^4$ \cmt,  such that H is mostly in the form of H$_2$ when the dense cloud begins to form around $t=33$ Myr. The transition between C$^+$, C, and CO occurs at $t\approx34.7$ Myr, where $n_\text{H} \approx 10^3$ \cmt. After this time, the carbon is mostly in the form of CO. As shown in Fig. \ref{fig:clump_11_stat}, after $t\approx35.4$ Myr the gas-phase CO significantly decreases and starts to deplete severely at the surface of the grains, where $A_\text{V} \gtrsim 7$ mag and $n_\text{H} \gtrsim 10^4$ \cmt.

\subsubsection{Spatial distribution of commonly observed gas-phase molecules}

\begin{figure*}
  \centering
  \includegraphics[width=15cm]{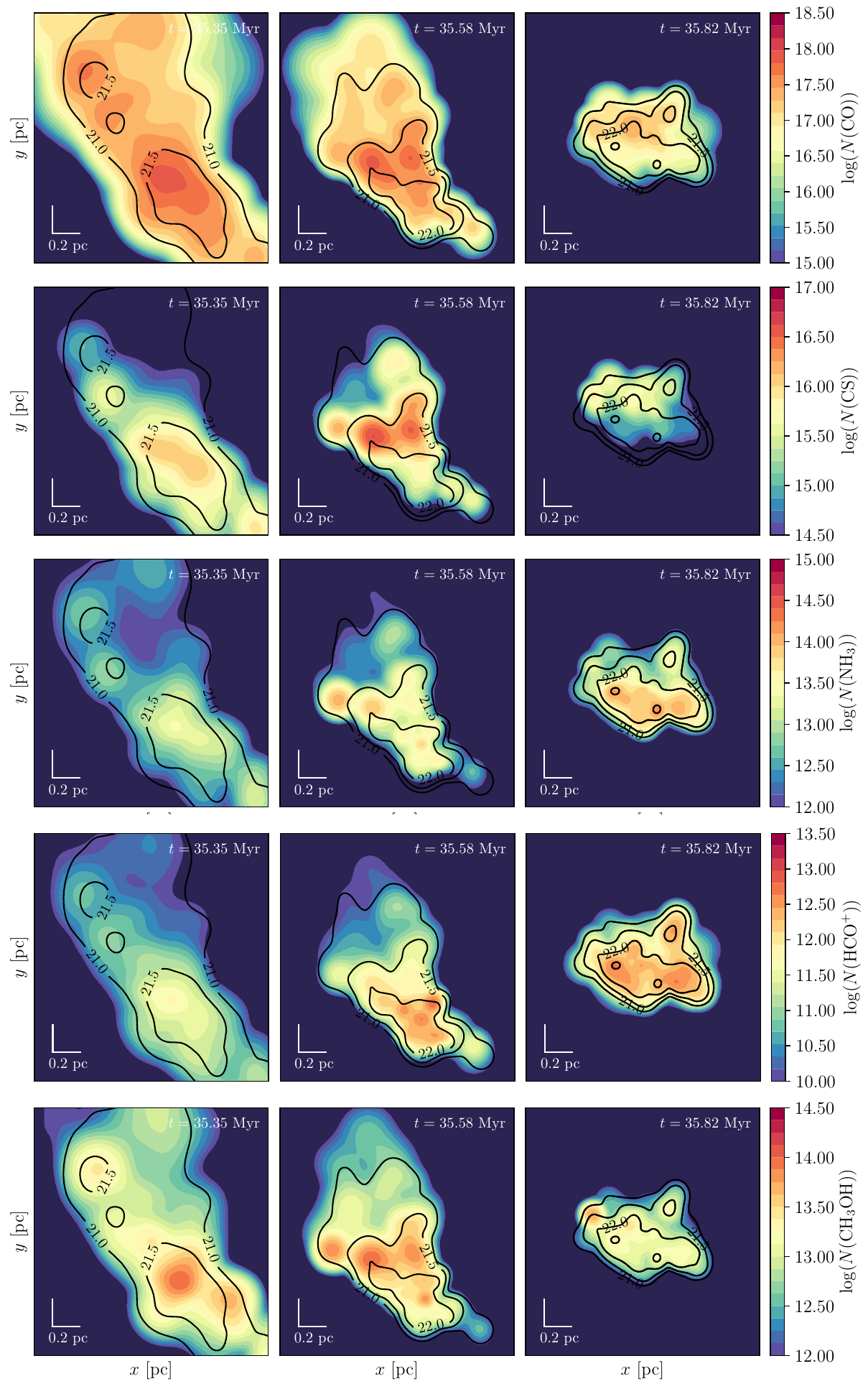}
    \caption{Computed gas-phase column density maps of CO, CS, NH$_3$, HCO$^+$, CH$_3$OH at $t=35.35$ Myr, $t=35.58$ Myr and $t=35.82$ Myr for cloud A. Contours were added for clarity and indicate $\log(N_\text{H})$ from 21 to 23.}
    \label{fig:clump_11_cdmap}
\end{figure*}

Fig. \ref{fig:clump_11_cdmap} presents the computed column density maps of CO, CS, NH$_3$, HCO$^+$, and CH$_3$OH at $t=35.35$ Myr, $t=35.58$ Myr, and $t=35.82$ Myr for cloud A. Contour lines were added for clarity and indicate $\log(N_\text{H})$ ranging from 21 to 23. These computed column densities are in general in good agreement with what is observed in dense clouds presenting similar total column density. At $t=35.35$ Myr most of the molecules shown in this figure follow the distribution of $N_\text{H}$. As discussed before, the gas-phase CO, as well as CS here, starts to severely deplete onto grains (this mechanism is also called the freezing out) around $t=35.40$ Myr, leading to a decrease of their gas-phase abundance. This freeze-out leads to an abundance gradient of these molecules along the north-south direction when the cloud has reached its peak density. On the contrary, NH$_3$ and CH$_3$OH are well distributed in the entire cloud and roughly follow the distribution of $N_\text{H}$. Such selective depletion is commonly observed in pre-stellar cores \citep[e.g.,][]{Willacy98,Caselli99,Bergin02,Bacmann02,Tafalla04,Lippok13}. Authors such as \citet{Willacy98}, \citet{Tafalla04}, or \citet{Pagani05} have shown that in several sources, molecules such as CO, CS, CCS, or SO are strongly depleted in core interiors, while species such as N$_2$H$^+$ and NH$_3$ show almost no signs of depletion. As suggested by previous models \citep[e.g.,][]{Bergin97} and observations, we find that the depletion of CO and CS is triggered by the local density of each SPH particle. The homogeneity of the spatial distribution of NH$_3$ and CH$_3$OH is found to be closely linked with their formation pathways, which involves the grain surface chemistry. For these two molecules we find that their formation on grain surfaces efficiently compete with their accretion on grains, which as for CO and CS tends to decrease their gas-phase abundance. However, in contrast to CO and CS, their efficient formation on grain surfaces allow their continuous replenishment of the gas phase, which explain the homogeneity of their spatial distribution.

\subsubsection{Grain surface chemistry and ice composition}
For this cloud, we found that the first monolayer of ice is formed around $t\approx34.5$ Myr, where $T_\text{dust}\approx 13$ K, $n_\text{H} \approx 10^3$ \cmt~ and $A_\text{V} \approx 1.3$ mag. This value is very close to the typical threshold observed in the interstellar medium $A_\text{V} \approx 1.6$ \citep{Whittet01,Boogert15}.  At this time, the low dust temperature, which ensures a relatively low rate of thermal desorption of most of the molecules present on the surface, and the significant shielding by dust, which ensures a low rate of photodesorption, allow molecules to stay physisorbed on grains and the mantle to grow. As suggested by the observations \citep[see][for a review on observations of interstellar ices]{Boogert15}, we found that the few first monolayers of ice are polar ices\footnote{In general the different type of ices, i.e., H$_2$O rich and CO rich, are referred to polar and apolar ices due to the different molecular dipole moments. Apolar ices are dominated by molecules with low dipole moments (e.g., CO, N$_2$) that have relatively low sublimation temperatures \citep{Boogert15}.} and that they mainly consist of H$_2$O ($\approx 80 \%$ of the total ice material), NH$_3$ ($\approx 11 \%$ of the total ice material), and CH$_4$ ($\approx 3 \%$ of the total ice material). At this time, these three molecules are found to be formed by successive hydrogenation of atomic oxygen\footnote{We note that, as already mentioned, for this cloud the transition from C$^+$ to CO occurs at $t\approx34.7$ Myr, such that most of the oxygen is mainly in its atomic form when the first monolayers of ice start to form.}, nitrogen, and carbon on grain surfaces, respectively.

\begin{figure*}
  \centering
  \includegraphics[width=15cm]{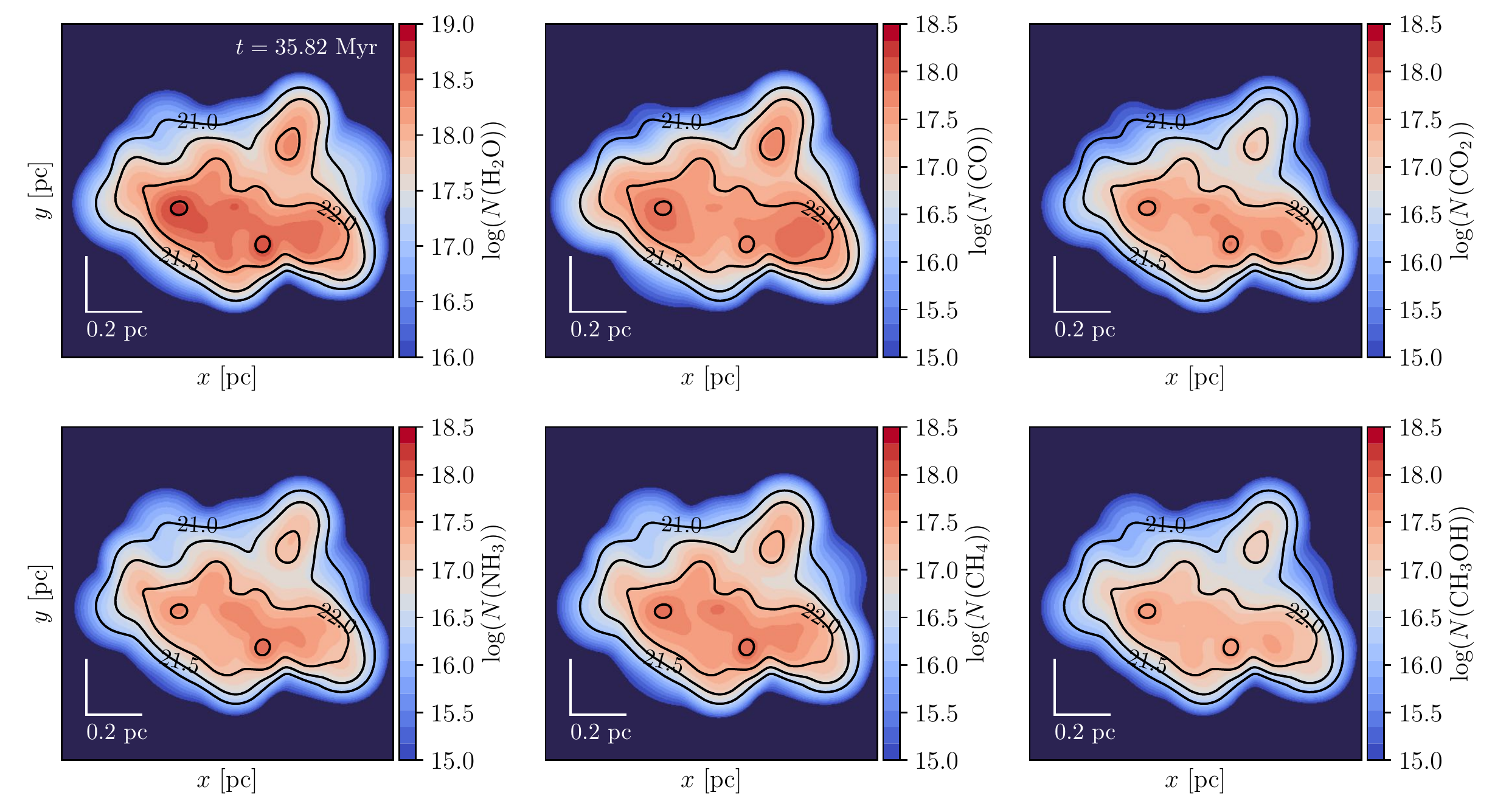}
    \caption{Computed column density map of H$_2$O, CO, CO$_2$, NH$_3$, CH$_4$, and CH$_3$OH on grains at the peak density of cloud A (at $t=35.82$ Myr). Contours were added for clarity and indicates $\log(N_\text{H})$ from 21 to 23.}
    \label{fig:clump_11_cdmap_carbon_grain}
\end{figure*}

At the peak density of the cloud, i.e., at $t\approx36$ Myr, $n_\text{H}\approx 4\times10^4$ \cmt, $T_\text{gas}\approx 12$ K, and $A_\text{V} \approx 10$ mag, a thickness of $\approx 100$ monolayers of ice is reached. H$_2$O is found to be the main constituent of the ice ($\approx40 \%$ of the total ice material). Other contributions come from CO ($\approx25 \%$ with respect to H$_2$O for its median abundance), CO$_2$, CH$_4$, HCOOH ($\approx10 \%$ each), NH$_3$, N$_2$, HS ($\approx8 \%$ each) H$_2$S, CH$_3$OH ($\approx6 \%$ each), CH$_2$NH$_2$, HCN, H$_2$CO ($\approx2 \%$ each), and CH$_3$O, HCO, ... ($\lesssim 2 \%$ each). The computed column densities of the main observed grain species at the peak density of the cloud are shown in Fig. \ref{fig:clump_11_cdmap_carbon_grain}.

\subsection{The case of one particular cloud: Cloud B}
\label{sec:pca_cloudB}

\begin{figure*}
  \centering
  \includegraphics[width=15cm]{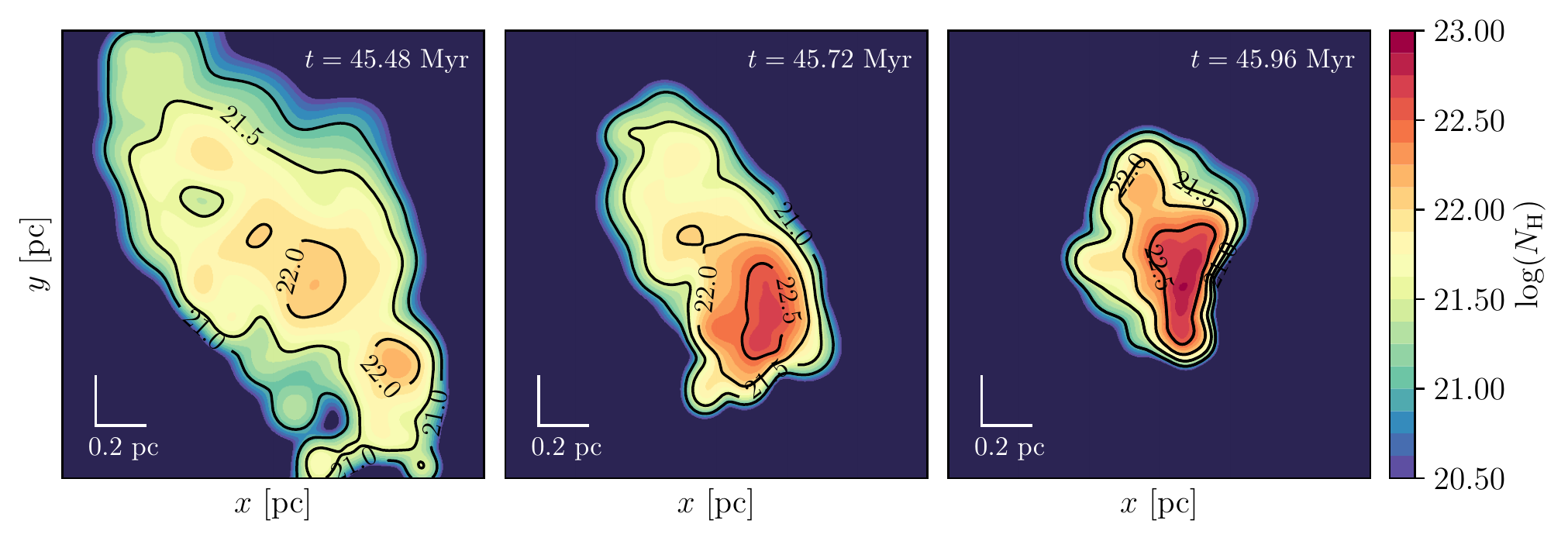}
    \caption{Evolution of the total column density map of cloud B. This figure shows the total column density of the cloud at $t=45.48$ Myr, $t=45.72$ Myr, and $t=45.96$ Myr, respectively. This cloud is formed by $\approx 300$ SPH particles.}
    \label{fig:clump_7_evol}
\end{figure*}

\begin{figure}
  \centering
  \includegraphics[width=9.2cm]{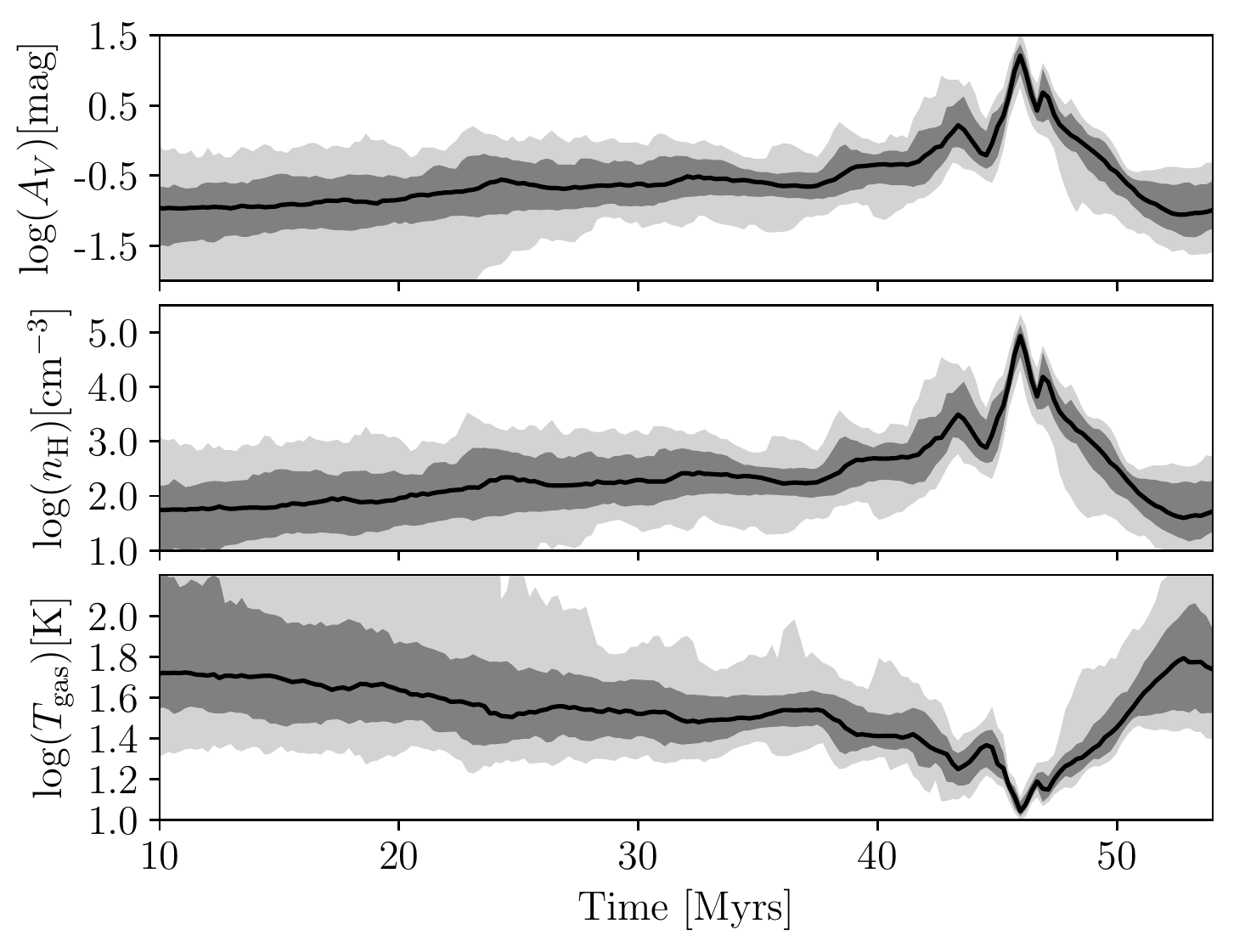}
    \caption{Statistics associated with the evolution of the physical parameters of all the SPH particles that compose cloud B. The black line shows the evolution of the median of the physical parameters. The dark gray filled area shows the 16th and 84th percentiles and the light gray filled area indicates the 2nd and 98th percentiles of the distribution.}
    \label{fig:clump_7_stat}
\end{figure}

\begin{figure*}
  \centering
  \includegraphics[width=15cm]{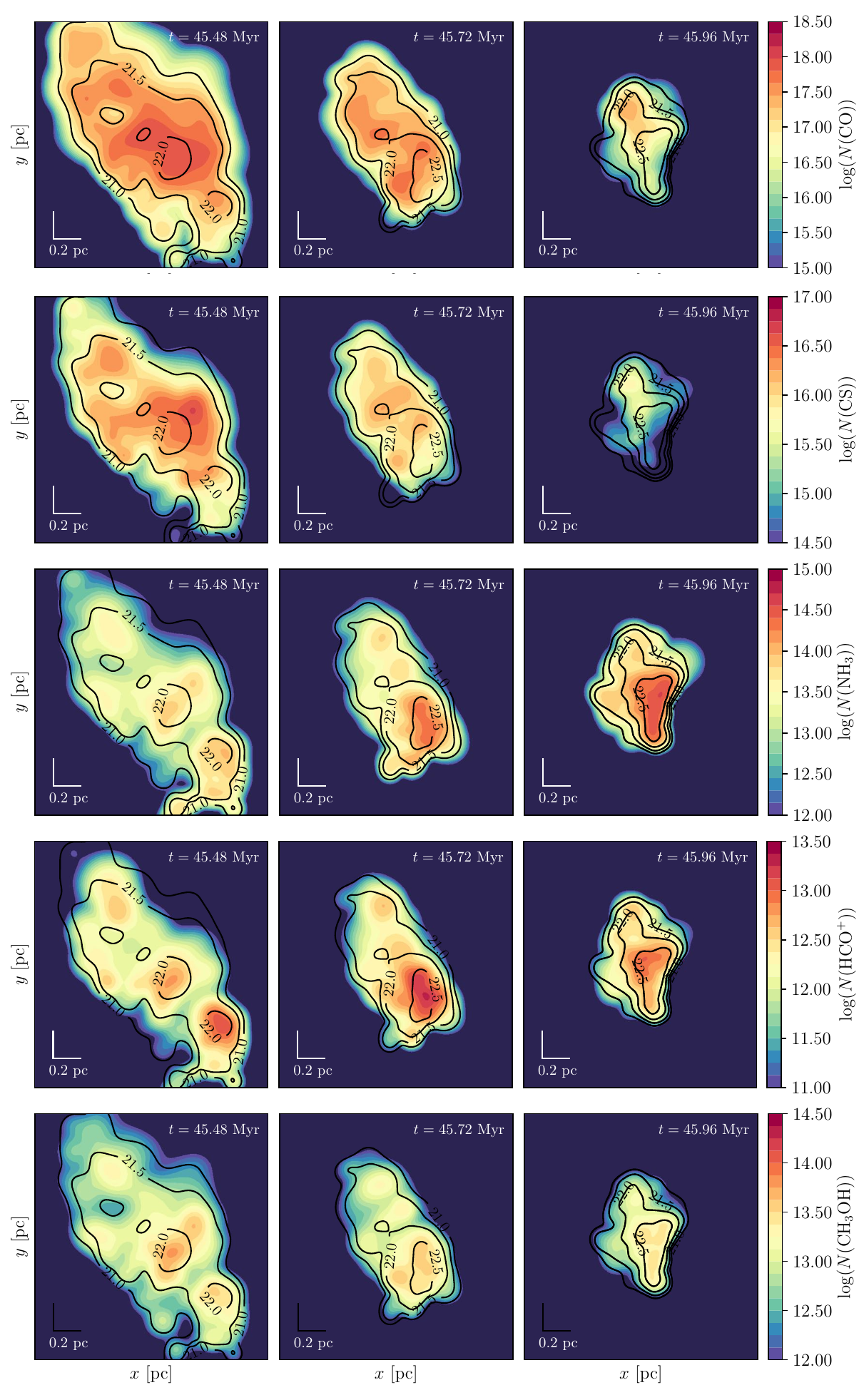}
    \caption{Same as Fig. \ref{fig:clump_11_cdmap} but for cloud B.}
    \label{fig:clump_7_cdmap}
\end{figure*}

In this section, we present results obtained for one particular cloud for which we observe strong abundance gradients within the cloud (denoted as cloud B). The evolution of the morphology of the cloud is shown in Fig. \ref{fig:clump_7_evol}, from $t=45.48$ Myr to $t=45.96$ Myr. The statistics associated with the evolution of the physical parameters of all the SPH particles that compose this cloud are shown in Fig. \ref{fig:clump_7_stat}.  As in the case of cloud A, all the SPH particles that compose this cloud have diffuse and translucent gas physical conditions in the phases that precede the formation of the cloud. Similar to the previous cloud, the H-H$_2$ transition occurs well before the dense cloud begins to form; i.e., $t\approx20$ Myr, and we have $f_{\textrm{H}_2}\approx 1$ in the dense cloud regime. Fig. \ref{fig:clump_7_cdmap} shows the computed column density maps of CO, CS, NH$_3$, HCO$^+$, and CH$_3$OH at $t=45.48$ Myr, $t=45.72$ Myr, and $t=45.96$ Myr. As in the case of cloud A we find that CO and CS start to severely deplete around $t\approx45.7$ Myr. This also leads to an abundance gradient of these molecules along the northeast--southwest direction in the cloud, while NH$_3$ and CH$_3$OH closely follow the distribution of $N_\text{H}$.

\begin{figure*}
  \centering
  \includegraphics[width=15cm]{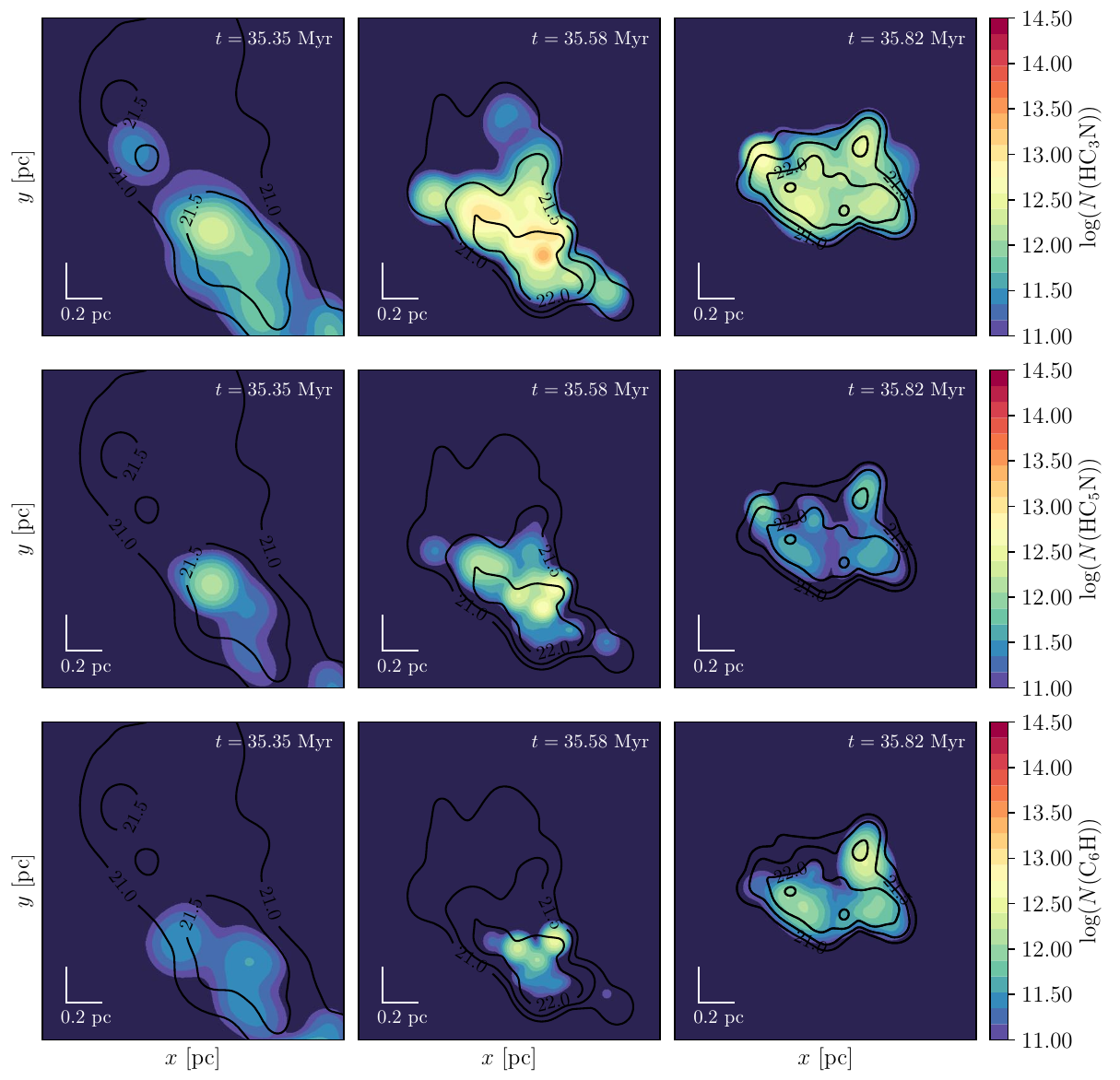}
  \caption{Computed gas-phase column density maps of HC$_3$N, HC$_5$N, and C$_6$H for cloud A.}
  \label{fig:clump_11_cdmap_carbon_chains}
\end{figure*}

\begin{figure*}
  \centering
  \includegraphics[width=15cm]{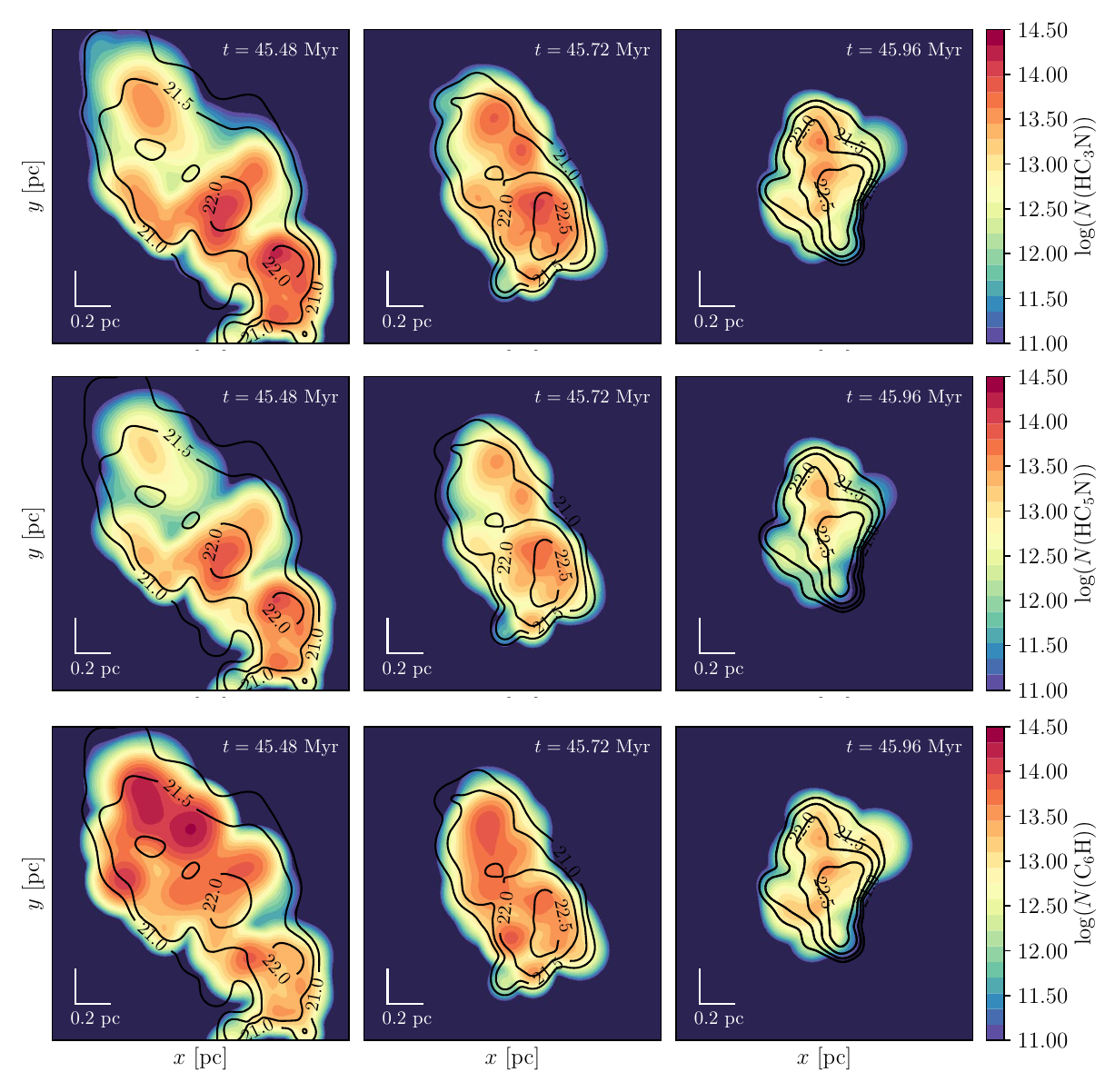}
 \caption{Same as Fig. \ref{fig:clump_11_cdmap_carbon_chains} but for cloud B.}
  \label{fig:clump_7_cdmap_carbon_chains}
\end{figure*}

Although the physical parameters in the densest region of this cloud are relatively similar to that of cloud A, we observed that the presence of an intermediate density pre-phase (i.e., for which \nh~ranges between few $10^2$ \cmt~and few $10^4$ \cmt) around $t\approx 43$ Myr strongly affects the predicted abundances of several molecules in the cold dense cloud regime. These differences are mostly found for carbon chains that are present at high column densities in cloud B. Figs. \ref{fig:clump_11_cdmap_carbon_chains} and \ref{fig:clump_7_cdmap_carbon_chains} present a comparison of the computed column density map of HC$_3$N, HC$_5$N, and C$_6$H for  clouds A and B. In addition to the fact that we find column densities of carbon chains $\sim 10$ times higher in cloud B than in cloud A, we also find that carbon chains show a relatively high dispersion in their abundances as compared to cloud A.

\begin{figure*}
  \centering
  \includegraphics[width=15cm]{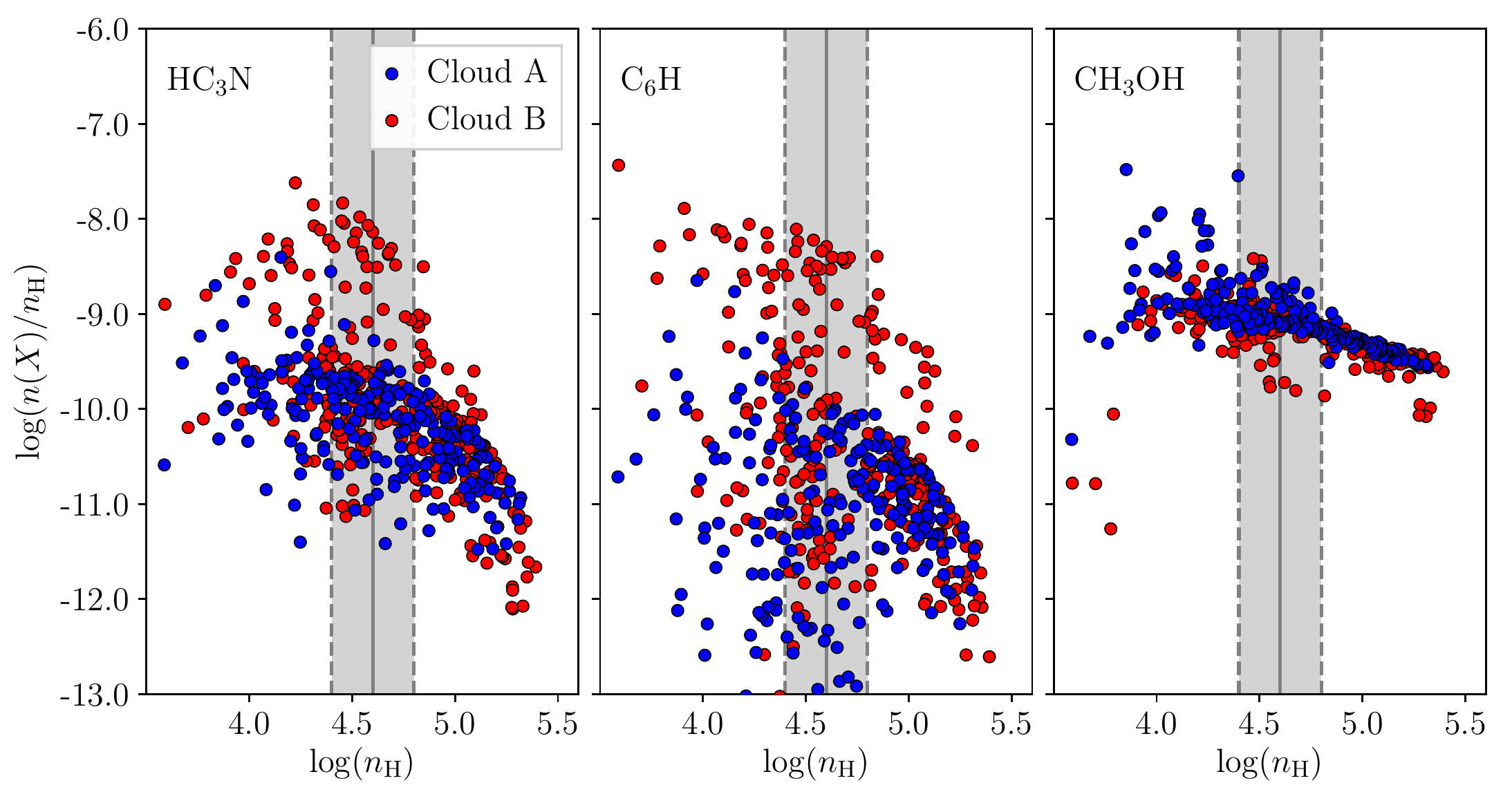}
    \caption{Computed gas-phase abundances of HC$_3$N, C$_6$H, CH$_3$OH as a function of the local density of each SPH particle at the peak density of cloud A and B, i.e., at $t=35.82$ Myr and $t=45.96$ Myr, respectively. Points in blue are the results obtained for cloud A, while points in red show the results obtained in the case of cloud B. The black line presents the median density of both clouds and the filled area indicates the range on which the abundance dispersion of Fig. \ref{fig:clump_abun} was computed.}
    \label{fig:clump_7_example}
\end{figure*}

\begin{figure}
  \centering
  \includegraphics[width=7.5cm]{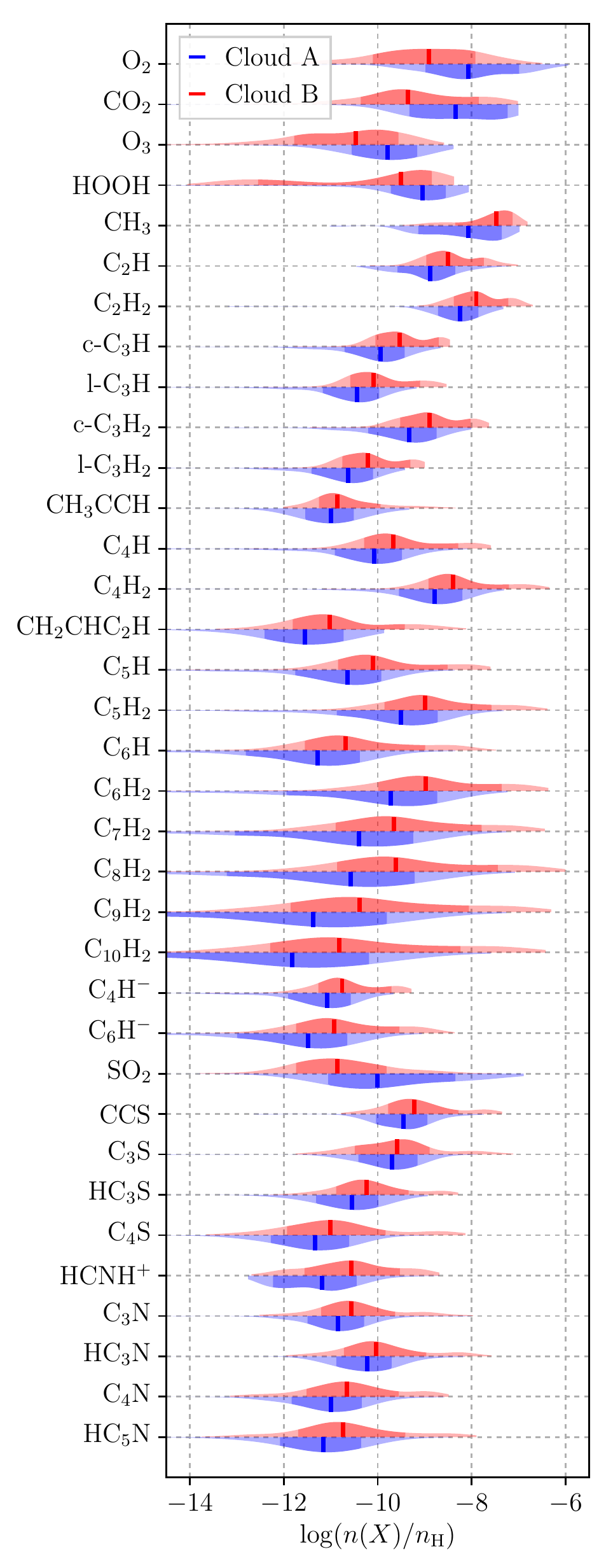}
    \caption{Distribution of all the gas-phase molecules with an abundance greater than $10^{-11}$ and presenting a difference greater than 1.5 dex on the abundances between the 16th and the 84th percentile in a range of 0.2 dex around $\log(n_\text{H}) = 4.6$, at the peak density of cloud A and B (as shown by the dark shaded area in Fig. \ref{fig:clump_7_example}). Points in blue are the results obtained for cloud A, while points in red show the results obtained in the case of cloud B.}
    \label{fig:clump_abun}
\end{figure}

This dispersion is illustrated in Fig. \ref{fig:clump_7_example}, which presents the abundance of three molecules, HC$_3$N, C$_6$H, and CH$_3$OH, as a function of the local density of each SPH particle at the peak density of each cloud. As shown in this figure, whilst for CH$_3$OH the results for both clouds are similar and show very little dispersion, the results obtained for HC$_3$N and C$_6$H show a significant dispersion (of at least 2 and 3 dex for HC$_3$N and C$_6$H, respectively at $n_\text{H}=4\times10^4$ \cmt). It is also important to note that this dispersion is larger in the case of cloud B, which shows enhanced abundances of these two molecules for some SPH particles; those particles with enhanced abundances are important because they are responsible for the high column density of these molecules. More generally, Fig. \ref{fig:clump_abun} presents all the gas-phase molecules with abundances greater than $10^{-11}$ that present a difference greater than 1.5 dex from the abundances between the 16th and 84th percentile in a range of 0.2 dex around $\log(n_\text{H}) = 4.6$, at the peak density (i.e., as shown by the dark shaded area in Fig. \ref{fig:clump_7_example}) of each cloud. The choice to restrain the density interval on which the dispersions are calculated is to try to remove the dependency on the local density of each particles and thus dispersion caused by the depletion on grains. The fact that this strong dispersion in the HC$_3$N and C$_6$H abundances (and more generally all the molecules presented in Fig. \ref{fig:clump_abun}) can be observed at a given density suggests that this dispersion is not only due to local physical conditions (e.g., adsorption on grain surface), but to the physical history that each particle underwent before reaching this time. As shown in Fig. \ref{fig:clump_abun}, most of these species are carbon chains. In this range of densities, the most important dispersion in the abundances is observed for complex carbon chains species that can be as high as four orders of magnitude for both clouds. 

From these results, at least three major questions arise: (1) What causes this dispersion? (2) Why are carbon chains the most affected? and (3) What explains the observed enhancement of carbon chains in cloud B as compared to cloud A? We try to answer these three questions in the following by focusing on the results obtained in cloud B.

\subsection{Link with physical history and local physical conditions}

In general, we find that this dispersion in the abundances observed in cloud B is strongly correlated to the electronic fraction of the medium,  gas-phase C/O ratio, and local density of each SPH particle.

\begin{figure*}
  \centering
  \includegraphics[width=15cm]{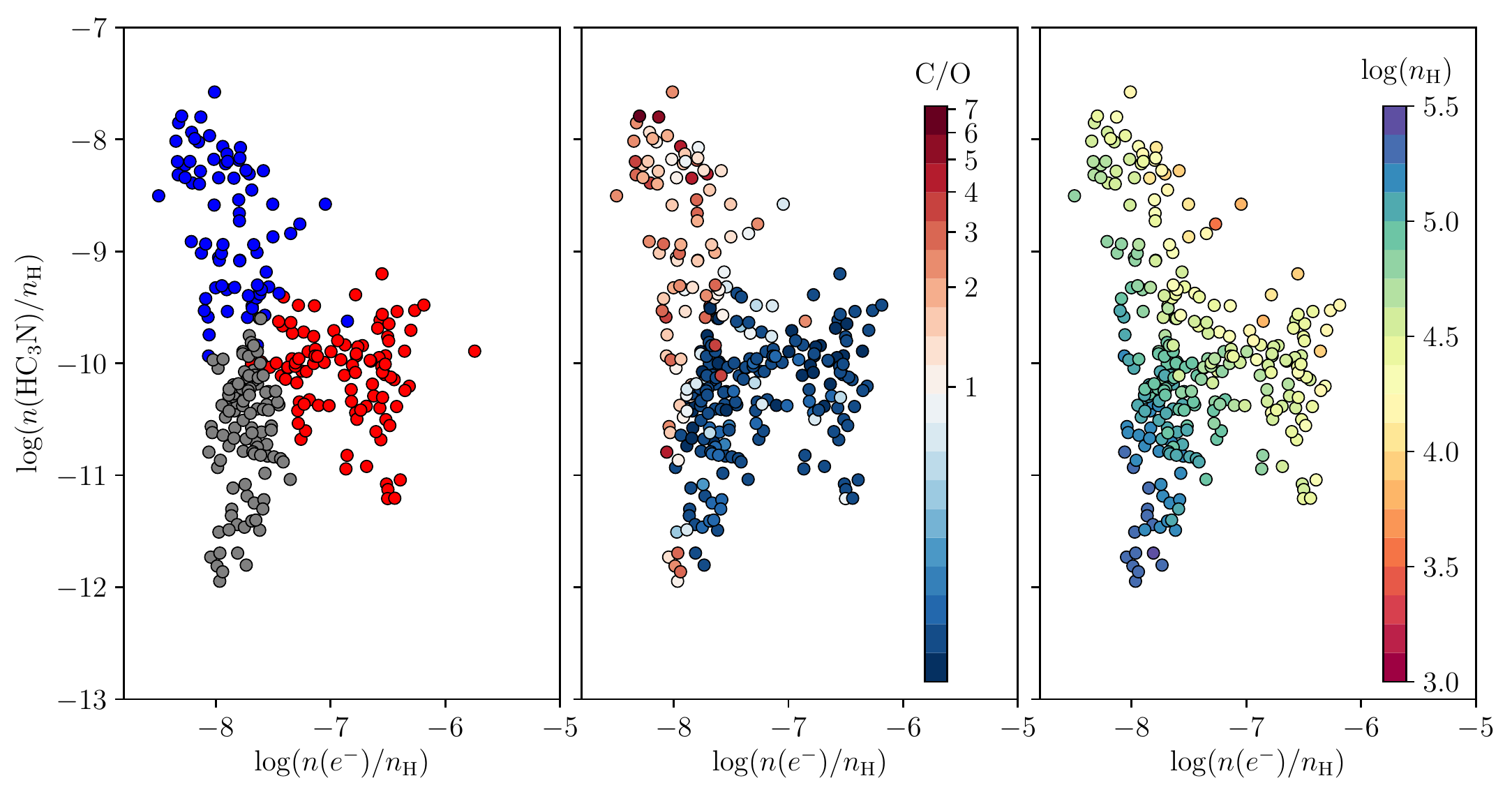}
    \caption{Computed abundances of HC$_3$N at $t=45.96$ Myr as a function of the local electronic fraction of each SPH particle at the peak density of cloud B. The left panel shows the results of the $k-$mean clustering using $k=3$ (the colors show the three identified clusters). This clustering was carried out on the PCs computed from the PCA analysis. The middle panel shows the same results but with a color coding that traces the C/O ratio computed just after the first density phase (i.e., at $t\approx 44.5$ Myr) and the right panel with a color coding that traces the individual density of each SPH particle. From these plots, we find that the cluster in blue mostly traces all the particles with a high abundance of HC$_3$N, a low electronic fraction on the order of $n(e^-)/n_\text{H} \approx 10^{-8}$, and a C/O ratio larger than 1. The cluster in red mostly traces all the particles with a moderate to low abundance of HC$_3$N, a high electronic fraction, and C/O $\lesssim1$. Finally, the cluster in gray traces all the particles with a low abundance of HC$_3$N, a low electronic fraction, and a high local density ($n_\text{H}\gtrsim 10^5$ \cmt).}
    \label{fig:clump_7_cluster}
\end{figure*}

To facilitate the interpretation, we clustered the results obtained with  cloud B using a $k$-means algorithm. In our approach, we wanted to cluster particles (i.e., $\sim300$ SPH particles), hence the abundances of each chemical species at each time step are the variables (i.e., $\sim900$ chemical species). In the $k$-means clustering algorithm, the number of clusters $k$ must be defined by the user and the correct choice of this parameter is often ambiguous. Given the size of the problem we first applied a principal component analysis \citep[PCA;][]{Jolliffe02} to reduce the dimensionality of the data set prior to clustering; a mean subtraction was performed over the data set before applying the PCA. This allowed us to validate the choice of the number of clusters that we used by making possible a direct visualization of the clustering results in the subspace of principal components (PCs). We then checked that the clustering results using PCs and original data give similar results. The clustering using PCs at $t\approx 46$ Myr was performed over the three first PCs, which contain more than $90\%$ of the variation in the modeling results ($\approx70\%$ for the first PC, $\approx 15\%$ for the second, and $\approx 6\%$ for the third). Fig. \ref{fig:clump_7_cluster} shows an example of the results obtained after clustering using PCs. These three plots present the abundance of HC$_3$N as a function of the electronic fraction of each SPH particle at the peak density of the cloud at $t \approx 46$ Myr. The left panel shows the result of the clustering obtained with $k=3$; the colors show the three clusters identified. The middle panel shows the same results but with a color coding that traces the C/O ratio computed just after the first density phase (i.e., at $t\approx 44.5$ Myr) and the right panel with a color coding that traces the individual density of each SPH particle at $t\approx 46$ Myr. It is important to note that the C/O ratio shown here does not correspond to the ratio between atomic C and O, but the ratio over all the carbon and oxygen molecular species present in the gas phase at this time.

As shown in Fig. \ref{fig:clump_7_cluster}, the first cluster in gray traces all the particles with a low abundance of HC$_3$N, a low electronic fraction and a high local density (typically $n_\text{H}\gtrsim 10^5$ \cmt).  In this case, the high density of each of these particles explains both the low abundance of HC$_3$N---and, more generally, all the molecules that do not have efficient grain surface formation routes that could allow their gas-phase replenishment---and the low electronic fraction (i.e., due to the accretion of the electron donors at the surface of the grains). 

The case of the two other clusters is interesting because even though they have roughly similar local physical conditions (as shown on the right panel of Fig. \ref{fig:clump_7_cluster} for the local density of each SPH particle), these two clusters show very different behaviors. Indeed, as shown in Fig. \ref{fig:clump_7_cluster}, the cluster in blue mostly traces all the particles with a high abundance of HC$_3$N (this is also the case for all the molecules presented in Fig. \ref{fig:clump_abun}), a low electronic fraction on the order of $n(e^-)/n_\text{H} \approx 10^{-8}$, and a C/O ratio larger than 1. On the contrary, the cluster in red mostly traces all the particles with a moderate to low abundance of HC$_3$N, a high electronic fraction, and C/O $\lesssim1$. In the case of this cluster, the abundance of HC$_3$N and the electronic fraction are highly correlated with the local density of the medium and decrease as the density increases.

\begin{figure}
  \centering
  \includegraphics[width=9.2cm]{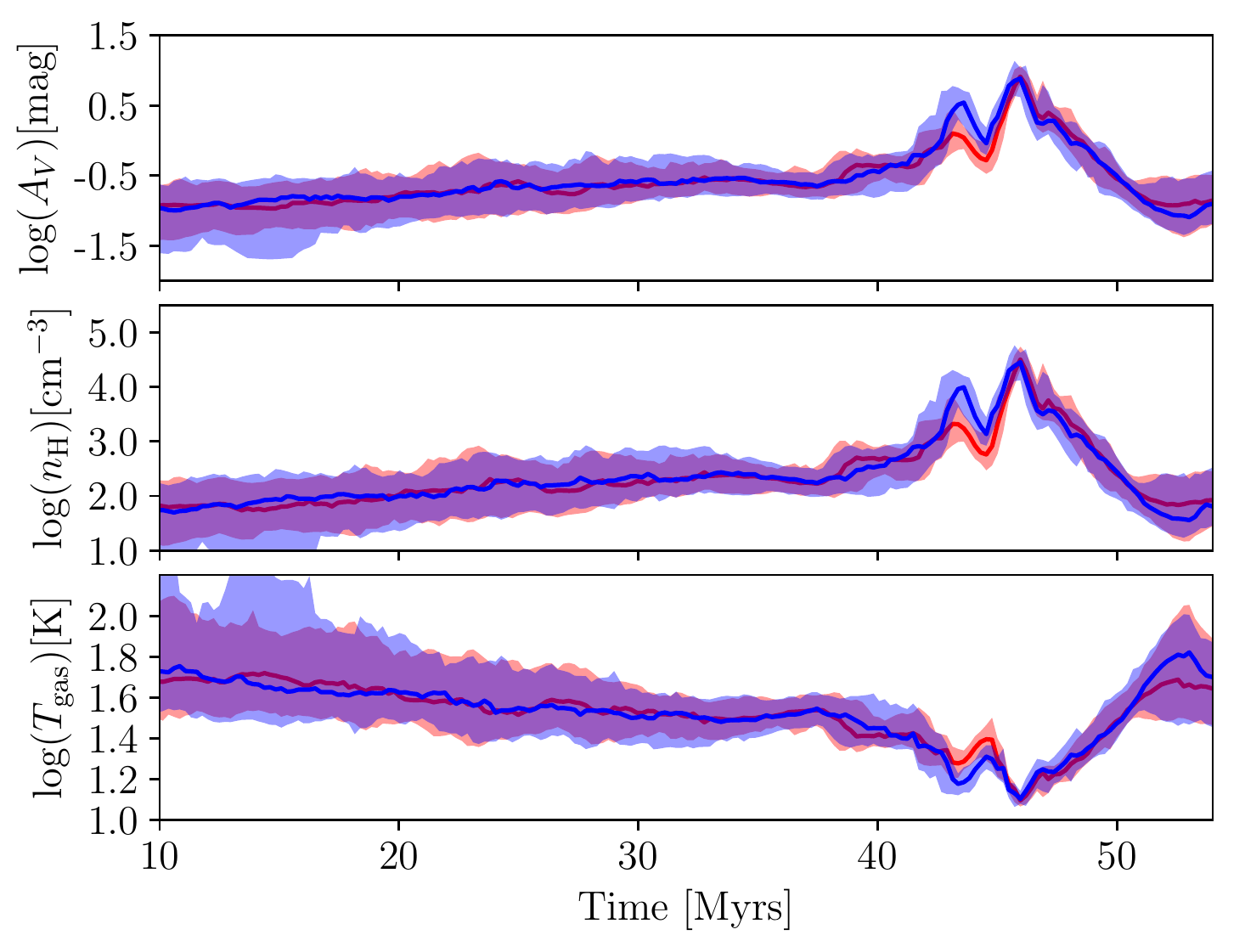}
    \caption{Same as Fig. \ref{fig:clump_7_stat} but with the results of the $k-$mean clustering presented in Fig. \ref{fig:clump_7_cluster} and Fig. \ref{fig:clump_7_metals}. All the SPH particles identified in the gray cluster in Figs. \ref{fig:clump_7_cluster} and \ref{fig:clump_7_metals} were removed for clarity.}
    \label{fig:clump_7_stat_clus}
\end{figure}

By tracing the history of each cluster  back in time, we find that the low electronic fraction and high C/O ratio observed in the cluster in blue are directly connected to the passage of these particles in the intermediate density pre-phase around $t=43$ Myr. This is illustrated in Fig. \ref{fig:clump_7_stat_clus}, which shows the statistics of the physical parameters associated with each particle cluster.

For this cluster, we find that, firstly, this intermediate density pre-phase reduces the abundance of the main electron donors due to their accretion on grains and thus reduces the electronic fraction of the cloud, which is now controlled by the cosmic ray ionization of H$_2$ and  He. This is shown in Fig. \ref{fig:clump_7_metals}, which shows the abundance of metals (which represents the sum over C$^+$, S$^+$, Si$^+$, Fe$^+$, Na$^+$, Mg$^+$, P$^+$, and Cl$^+$) as a function of the electronic fraction of each particle cluster at the peak density of cloud (i.e., at $t\approx$46 Myr).
 Secondly, this intermediate density pre-phase increases the gas-phase C/O ratio because of the formation of water ice on grain, which leads to a sequestration of atomic oxygen on grains. For this cluster, we find that almost 80 monolayers consisting of $\sim 80\%$ of water ice are present on the surface of the grains after this density phase. This means that for these SPH particles, a large amount of the elemental oxygen has been removed from the gas phase and included in the solid phase during this event.

These two effects have a strong impact on the formation of complex carbon chains. Indeed, carbon chains are mainly formed by ionic pathways and neutral-neutral reactions in which insertion of carbon atoms is a crucial step. 

Therefore, on the one hand, by reacting with intermediate ions, electrons play an important role in setting this synthetic reaction chain mechanism efficiency \citep[see][for a complete review of the chemistry in dark clouds]{Agundez13}. Removing the main electron donors from the gas phase have the consequence of promoting the ion-neutral chemistry in the sense that intermediates are less destroyed by electronic recombination. 

\begin{figure*}
  \centering
  \includegraphics[width=15cm]{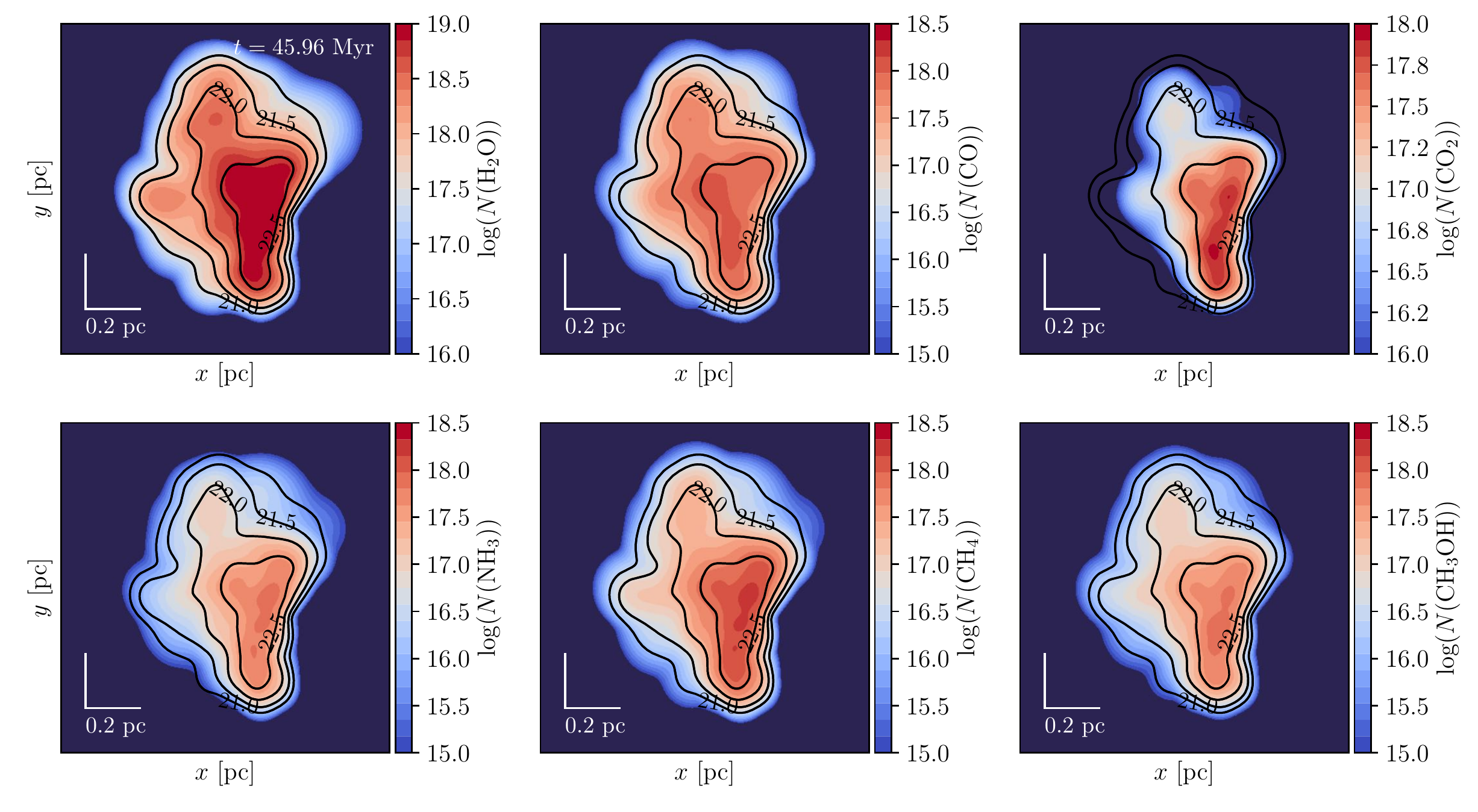}
    \caption{Similar to Fig. \ref{fig:clump_11_cdmap_carbon_grain} but for cloud B.}
    \label{fig:clump_7_cdmap_carbon_grain}
\end{figure*}

\begin{figure}
\centering
  \includegraphics[width=7cm]{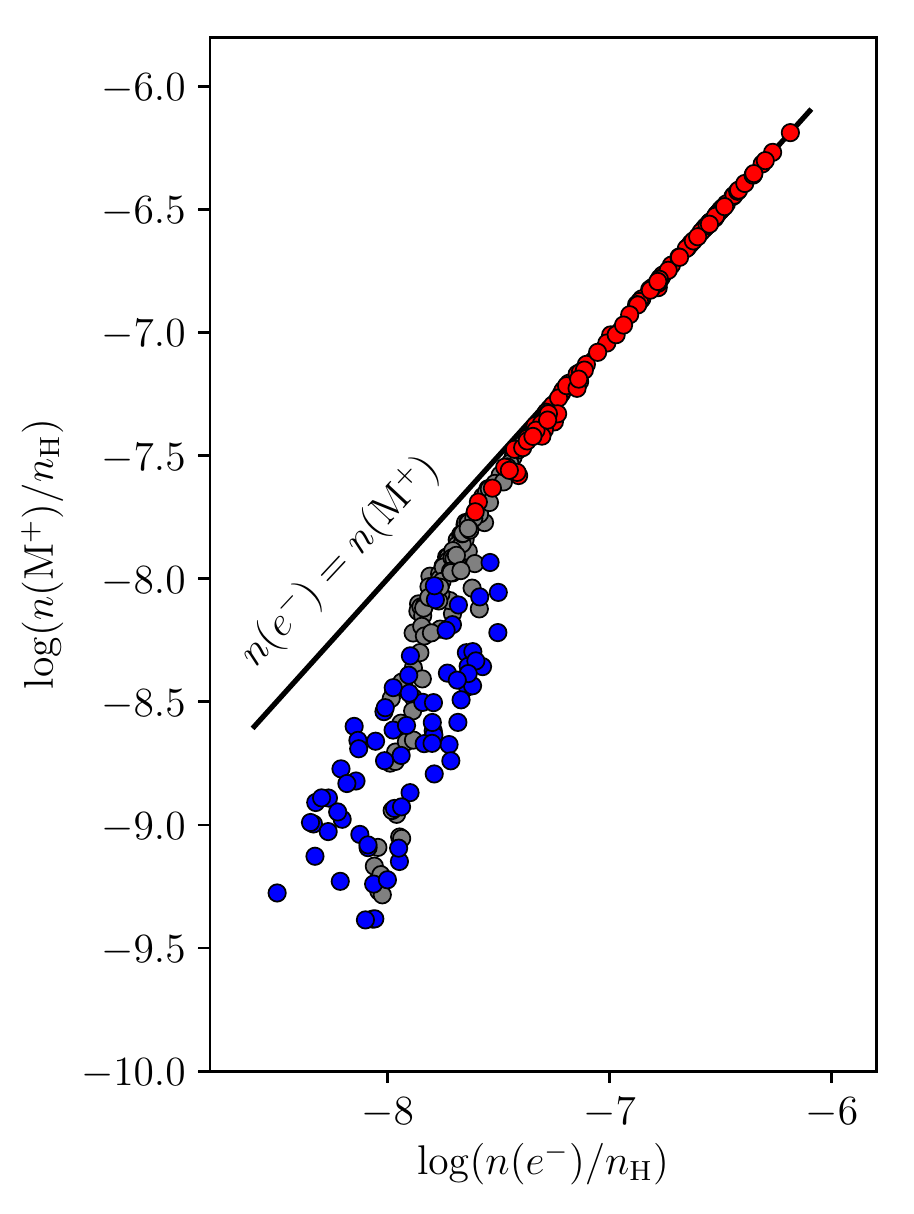}
  \caption{Computed abundance of metals (i.e., sum over C$^+$, S$^+$, Si$^+$, Fe$^+$, Na$^+$, Mg$^+$, P$^+$, and Cl$^+$) at $t=45.96$ Myr as a function of the electronic fraction of each particle cluster at the peak density of cloud B. The same color coding as on the left panel of Fig. \ref{fig:clump_7_cluster} is used.}
  \label{fig:clump_7_metals}
\end{figure}

On the other hand, the formation water ice on grains increases the amount of free carbon, which is otherwise mostly locked up into CO, that can be used for the synthesis of carbon chains. However, as shown in the middle panel of Fig.  \ref{fig:clump_7_cluster}, a higher C/O ratio does not necessarily mean a higher abundance of complex carbon chains. This strongly depends on the amount of atomic and ionized carbon available and thus on the local physical conditions of each SPH particle. For this cluster, we find that this water rich ice survives until the cloud forms. However, when looking at the spatial distribution of molecules in the cloud, we find that this does not produce observable constraints that could allow us to trace such history. This is illustrated in Figs. \ref{fig:clump_11_cdmap_carbon_grain} and \ref{fig:clump_7_cdmap_carbon_grain}, which present the computed column densities of the main observed grain species at the peak density of clouds A and B, respectively.

On the contrary, most of the particles in cluster red have not experienced such a high density pre-phase (or at least with a lower intensity). This results in an lower C/O ratio, meaning a lower abundance of free carbon, which is mostly locked up into CO, but also, as shown in Fig. \ref{fig:clump_7_metals}, a higher abundance of metals. This implies a relatively high electronic fraction, which prevents the efficient formation of carbon chains.

This link with the past history of each SPH particle also explains the observed deficit of carbon chains in cloud A as compared to cloud B. Indeed, the fact that cloud A evolves continuously from the diffuse medium (i.e., without density pre-phase) to the dense cloud regime leads to a high electronic fraction ($n(e-)/n_\text{H}\approx 2\times10^{-7}$ at $t=35.82$ Myr). As a consequence, by reacting with intermediate ions, this high abundance of electrons prevents the formation of complex carbon chains in the cloud.

\section{Discussion and analogy with observations of dense clouds}
\label{sec:discussion}

Several attempts have been made to explain the unusual high abundances of carbon chains observed in the TMC-1 dense cloud. One of the most popular scenarios relies on the so-called early time chemistry. This scenario is based on the fact that carbon chain chemistry requires carbon to be present in a reactive form in the gas phase. In pseudo-time-dependent models, this condition is satisfied at early time, i.e., before few $10^5$ years after which the carbon is locked up in CO \citep[see Fig. 3 of][for exemple]{Agundez13}. In this scenario, molecular clouds are thus considered to be seen in an early evolutionary stage in which the available carbon is still in an reactive form. Another possible explanation has been to invoke a different C/O ratio in TMC-1 as compared to other clouds \citep{Bergin96}. In both case, the net effect of these scenarios is to increase the amount of free carbon that can be used for the synthesis of these molecules. 

Some authors have proposed a lower electronic fraction of TMC-1 as a good explanation of these unusual abundances. As discussed earlier, the effect of the electronic fraction on the synthesis of complex molecules has been early recognized by \citet{Graedel82} who proposed so-called low metal abundances. However, it has been shown that the use of these low metal abundances could not itself explain the observed abundances of carbon chains in this cloud \citep[see][for exemple]{Agundez13}. Several others studies have tried to find some explanations for a possible lower electronic fraction in TMC-1 as compared to others clouds. \citet{Flower07}  for example showed that a lower cosmic ray ionization rate in TMC-1 could help explain the carbon chain abundances of this cloud, while \citet{Wakelam08} discussed the importance of PAHs on their synthesis.

In this study we propose that the observed differences between dense clouds may arise from different dynamical evolution. We show that both the local physical conditions and the history of each SPH particle have an impact on the computed abundances of each dense clouds. We find that these differences mainly arise from (1) the electronic fraction, (2) the C/O ratio, and (3) the local physical conditions (i.e., the local density in our case). Although these mechanisms have already been invoked by previous studies, we show their direct link with the dynamics at large scale. This gives a more natural explanation of the observed differences between dense clouds but also within some of them. In particular, these results show that the often invoked early time chemistry is not a requisite to produce high abundances of carbon chains. On the contrary, based on the results obtained with our cloud B, we show that high abundance of carbon chains can be the result of more evolved gas, i.e., having experienced a past density phase in this case.

In this context and by analogy with the observations, we argue that the strong abundance gradients observed on sub-parsec scales in the TMC-1 dense cloud \citep{Olano88,Pratap97} may originate from a different dynamical evolution of the gas that formed it. In particular, the observed enhancement of carbon chains in the so-called cyanopolyyne peak may, as in the case of our cloud B, traces a past density phase in which most of the electron donors have been removed from the gas phase, enhancing the ion-molecule chemistry. On the same basis, it could also explain the low abundance of carbon chains in the L134N dense cloud, which in this case would have a higher electronic fraction than TMC-1 and may be closer to our cloud A. 

More generally, our simulations suggest that the observed diversity in the molecular composition of dense clouds, and especially complex carbon chains, is strongly linked to the dynamical evolution of the interstellar matter. However, this study suffers from some limitations that should be addressed in the future. First, the fact that these simulations do not include the effect of small scale self-gravity could affect the morphologies of the considered clouds and thus their physical structure. Nevertheless, we think that including this mechanism leaves the overall conclusion of this paper unchanged, since the main observed differences between cloud A and B are found to be mainly driven by the effect of large scale dynamics. A major limitation could also come from the fact that the hydrodynamical simulations that we have used do not consider any localized sources of stellar feedback which could, for example, locally enhance the radiation flux. In this case, this could have a strong impact on the modeled results,  especially for cloud B. In this particular case, the presence of a localized source of radiation could be important enough to remove material from the surface of the grains after the intermediate density pre-phase (either by direct photodesorption or thermal evaporation of the ice if the temperature gets high enough) and thus significantly affect the results. In this case a proper modeling of the entire region would be required to track the evolution of the surrounding gas and thus the possibility to form stars in the vicinity of the considered clouds.

\section{Conclusions}
\label{sec:conclusion}

In this study we used results from large scale hydrodynamical simulations to study the transition between diffuse to dense clouds on the molecular composition of the dense interstellar medium. We used results from SPH simulations by \citet{Bonnell13} from which we extracted physical parameters as a function of time that have been used as inputs for our full gas-grain chemical model.
Our results show that the molecular composition of dense clouds show a high dispersion in the computed abundances and especially for carbon chains. We showed that besides the abundance gradient caused by the depletion on grains (triggered by the local density of each SPH particle), the history of each SPH particle has a strong impact on the computed abundances of carbon chains. Based on the results obtained for one particular cloud for which high column densities of carbon chains are found, we show that these differences arise from differences in (1) the electronic fraction (2) the C/O ratio, and (3) the local physical conditions (i.e., the local density of each SPH particle in our case). These differences are directly linked to a different dynamical evolution of the gas that formed the cloud for which some SPH particles went to an intermediate density pre-phase prior its formation. This intermediate density pre-phase reduces the electronic fraction of the cloud, but also affects the C/O ratio which becomes larger than 1 due to the formation of water at the surface of grains. Based on these results, we propose that the observed differences between dense clouds may arise from different dynamical evolution of the gas that formed these clouds.

\begin{acknowledgements}
This work has been founded by the European Research Council (Starting Grant 3DICE, grant agreement 336474). The authors are also grateful to the CNRS program "Physique et Chimie du Milieu Interstellaire" (PCMI) co-funded by the Centre National d'Etudes Spatiales (CNES) for partial funding of their work. IAB gratefully acknowledges support from the ECOGAL project, grant agreement 291227, funded by the European Research Council under ERC-2011-ADG. This work relied on the computing resources of the DiRAC Complexity system, operated by the University of Leicester IT Services, which forms part of the STFC DiRAC HPC facility (www.dirac.ac.uk).
\end{acknowledgements}

\bibliographystyle{bibtex/aa}
\bibliography{bibtex/bibliography}

\begin{thebibliography}{56}
\expandafter\ifx\csname natexlab\endcsname\relax\def\natexlab#1{#1}\fi

\bibitem[{{Ag{\'u}ndez} \& {Wakelam}(2013)}]{Agundez13}
{Ag{\'u}ndez}, M. \& {Wakelam}, V. 2013, Chemical Reviews, 113, 8710

\bibitem[{{Bacmann} {et~al.}(2002){Bacmann}, {Lefloch}, {Ceccarelli},
  {Castets}, {Steinacker}, \& {Loinard}}]{Bacmann02}
{Bacmann}, A., {Lefloch}, B., {Ceccarelli}, C., {et~al.} 2002, \aap, 389, L6

\bibitem[{{Bergin} {et~al.}(2002){Bergin}, {Alves}, {Huard}, \&
  {Lada}}]{Bergin02}
{Bergin}, E.~A., {Alves}, J., {Huard}, T., \& {Lada}, C.~J. 2002, \apjl, 570,
  L101

\bibitem[{{Bergin} {et~al.}(2004){Bergin}, {Hartmann}, {Raymond}, \&
  {Ballesteros-Paredes}}]{Bergin04}
{Bergin}, E.~A., {Hartmann}, L.~W., {Raymond}, J.~C., \& {Ballesteros-Paredes},
  J. 2004, \apj, 612, 921

\bibitem[{{Bergin} \& {Langer}(1997)}]{Bergin97}
{Bergin}, E.~A. \& {Langer}, W.~D. 1997, \apj, 486, 316

\bibitem[{{Bergin} {et~al.}(1996){Bergin}, {Snell}, \& {Goldsmith}}]{Bergin96}
{Bergin}, E.~A., {Snell}, R.~L., \& {Goldsmith}, P.~F. 1996, \apj, 460, 343

\bibitem[{{Bergin} \& {Tafalla}(2007)}]{Bergin07}
{Bergin}, E.~A. \& {Tafalla}, M. 2007, \araa, 45, 339

\bibitem[{{Bertoldi} \& {McKee}(1992)}]{Bertoldi92}
{Bertoldi}, F. \& {McKee}, C.~F. 1992, \apj, 395, 140

\bibitem[{{Bohlin} {et~al.}(1978){Bohlin}, {Savage}, \& {Drake}}]{Bohlin78}
{Bohlin}, R.~C., {Savage}, B.~D., \& {Drake}, J.~F. 1978, \apj, 224, 132

\bibitem[{{Bonnell} {et~al.}(2013){Bonnell}, {Dobbs}, \& {Smith}}]{Bonnell13}
{Bonnell}, I.~A., {Dobbs}, C.~L., \& {Smith}, R.~J. 2013, \mnras, 430, 1790

\bibitem[{{Boogert} {et~al.}(2015){Boogert}, {Gerakines}, \&
  {Whittet}}]{Boogert15}
{Boogert}, A.~C.~A., {Gerakines}, P.~A., \& {Whittet}, D.~C.~B. 2015, \araa,
  53, 541

\bibitem[{{Caselli} {et~al.}(1999){Caselli}, {Walmsley}, {Tafalla}, {Dore}, \&
  {Myers}}]{Caselli99}
{Caselli}, P., {Walmsley}, C.~M., {Tafalla}, M., {Dore}, L., \& {Myers}, P.~C.
  1999, \apjl, 523, L165

\bibitem[{{Clark} {et~al.}(2012){Clark}, {Glover}, {Klessen}, \&
  {Bonnell}}]{Clark12}
{Clark}, P.~C., {Glover}, S.~C.~O., {Klessen}, R.~S., \& {Bonnell}, I.~A. 2012,
  \mnras, 424, 2599

\bibitem[{{Dickens} {et~al.}(2000){Dickens}, {Irvine}, {Snell}, {Bergin},
  {Schloerb}, {Pratap}, \& {Miralles}}]{Dickens00}
{Dickens}, J.~E., {Irvine}, W.~M., {Snell}, R.~L., {et~al.} 2000, \apj, 542,
  870

\bibitem[{{Dobbs} {et~al.}(2014){Dobbs}, {Krumholz}, {Ballesteros-Paredes},
  {Bolatto}, {Fukui}, {Heyer}, {Low}, {Ostriker}, \&
  {V{\'a}zquez-Semadeni}}]{Dobbs14}
{Dobbs}, C.~L., {Krumholz}, M.~R., {Ballesteros-Paredes}, J., {et~al.} 2014,
  Protostars and Planets VI, 3

\bibitem[{{Draine} \& {Lee}(1984)}]{Draine84}
{Draine}, B.~T. \& {Lee}, H.~M. 1984, \apj, 285, 89

\bibitem[{{Draine} \& {Li}(2007)}]{Draine07}
{Draine}, B.~T. \& {Li}, A. 2007, \apj, 657, 810

\bibitem[{{Flower} {et~al.}(2007){Flower}, {Pineau Des For{\^e}ts}, \&
  {Walmsley}}]{Flower07}
{Flower}, D.~R., {Pineau Des For{\^e}ts}, G., \& {Walmsley}, C.~M. 2007, \aap,
  474, 923

\bibitem[{{Gerin} {et~al.}(2003){Gerin}, {Foss{\'e}}, \& {Roueff}}]{Gerin03}
{Gerin}, M., {Foss{\'e}}, D., \& {Roueff}, E. 2003, in SFChem 2002: Chemistry
  as a Diagnostic of Star Formation, ed. C.~L. {Curry} \& M.~{Fich}, 81

\bibitem[{{Gerin} {et~al.}(2016){Gerin}, {Neufeld}, \& {Goicoechea}}]{Gerin16}
{Gerin}, M., {Neufeld}, D.~A., \& {Goicoechea}, J.~R. 2016, \araa, 54, 181

\bibitem[{{Glover} {et~al.}(2010){Glover}, {Federrath}, {Mac Low}, \&
  {Klessen}}]{Glover10}
{Glover}, S.~C.~O., {Federrath}, C., {Mac Low}, M.-M., \& {Klessen}, R.~S.
  2010, \mnras, 404, 2

\bibitem[{{Graedel} {et~al.}(1982){Graedel}, {Langer}, \&
  {Frerking}}]{Graedel82}
{Graedel}, T.~E., {Langer}, W.~D., \& {Frerking}, M.~A. 1982, \apjs, 48, 321

\bibitem[{{Hassel} {et~al.}(2010){Hassel}, {Herbst}, \& {Bergin}}]{Hassel10}
{Hassel}, G.~E., {Herbst}, E., \& {Bergin}, E.~A. 2010, \aap, 515, A66

\bibitem[{{Hennebelle} \& {Falgarone}(2012)}]{Hennebelle12}
{Hennebelle}, P. \& {Falgarone}, E. 2012, \aapr, 20, 55

\bibitem[{{Herbst} \& {Klemperer}(1973)}]{Herbst73}
{Herbst}, E. \& {Klemperer}, W. 1973, \apj, 185, 505

\bibitem[{{Herbst} \& {Leung}(1989)}]{Herbst89}
{Herbst}, E. \& {Leung}, C.~M. 1989, \apjs, 69, 271

\bibitem[{{Hickson} {et~al.}(2016){Hickson}, {Wakelam}, \&
  {Loison}}]{Hickson16}
{Hickson}, K.~M., {Wakelam}, V., \& {Loison}, J.-C. 2016, Molecular
  Astrophysics, 3, 1

\bibitem[{{Hincelin} {et~al.}(2013){Hincelin}, {Wakelam}, {Commer{\c c}on},
  {Hersant}, \& {Guilloteau}}]{Hincelin13}
{Hincelin}, U., {Wakelam}, V., {Commer{\c c}on}, B., {Hersant}, F., \&
  {Guilloteau}, S. 2013, \apj, 775, 44

\bibitem[{{Hollenbach} {et~al.}(1991){Hollenbach}, {Takahashi}, \&
  {Tielens}}]{Hollenbach91}
{Hollenbach}, D.~J., {Takahashi}, T., \& {Tielens}, A.~G.~G.~M. 1991, \apj,
  377, 192

\bibitem[{{Jenkins}(2009)}]{Jenkins09}
{Jenkins}, E.~B. 2009, \apj, 700, 1299

\bibitem[{{Jolliffe}(2002)}]{Jolliffe02}
{Jolliffe}, I.~T. 2002, {Principal Component Analysis} (Springer: New York)

\bibitem[{{Klessen} \& {Glover}(2016)}]{Klessen16}
{Klessen}, R.~S. \& {Glover}, S.~C.~O. 2016, Star Formation in Galaxy
  Evolution: Connecting Numerical Models to Reality, Saas-Fee Advanced Course,
  Volume 43.~ISBN 978-3-662-47889-9.~Springer-Verlag Berlin Heidelberg, 2016,
  p.~85, 43, 85

\bibitem[{{Leung} {et~al.}(1984){Leung}, {Herbst}, \& {Huebner}}]{Leung84}
{Leung}, C.~M., {Herbst}, E., \& {Huebner}, W.~F. 1984, \apjs, 56, 231

\bibitem[{{Lippok} {et~al.}(2013){Lippok}, {Launhardt}, {Semenov}, {Stutz},
  {Balog}, {Henning}, {Krause}, {Linz}, {Nielbock}, {Pavlyuchenkov},
  {Schmalzl}, {Schmiedeke}, \& {Bieging}}]{Lippok13}
{Lippok}, N., {Launhardt}, R., {Semenov}, D., {et~al.} 2013, \aap, 560, A41

\bibitem[{{Liszt}(2009)}]{Liszt09}
{Liszt}, H.~S. 2009, \aap, 508, 783

\bibitem[{{Loison} {et~al.}(2016){Loison}, {Ag{\'u}ndez}, {Marcelino},
  {Wakelam}, {Hickson}, {Cernicharo}, {Gerin}, {Roueff}, \&
  {Gu{\'e}lin}}]{Loison16}
{Loison}, J.-C., {Ag{\'u}ndez}, M., {Marcelino}, N., {et~al.} 2016, \mnras,
  456, 4101

\bibitem[{{Mathis} {et~al.}(1983){Mathis}, {Mezger}, \& {Panagia}}]{Mathis83}
{Mathis}, J.~S., {Mezger}, P.~G., \& {Panagia}, N. 1983, \aap, 128, 212

\bibitem[{{Olano} {et~al.}(1988){Olano}, {Walmsley}, \& {Wilson}}]{Olano88}
{Olano}, C.~A., {Walmsley}, C.~M., \& {Wilson}, T.~L. 1988, \aap, 196, 194

\bibitem[{{Pagani} {et~al.}(2004){Pagani}, {Bacmann}, {Motte}, {Cambr{\'e}sy},
  {Fich}, {Lagache}, {Miville-Desch{\^e}nes}, {Pardo}, \& {Apponi}}]{Pagani04}
{Pagani}, L., {Bacmann}, A., {Motte}, F., {et~al.} 2004, \aap, 417, 605

\bibitem[{{Pagani} {et~al.}(2003){Pagani}, {Lagache}, {Bacmann}, {Motte},
  {Cambr{\'e}sy}, {Fich}, {Teyssier}, {Miville-Desch{\^e}nes}, {Pardo},
  {Apponi}, \& {Stepnik}}]{Pagani03}
{Pagani}, L., {Lagache}, G., {Bacmann}, A., {et~al.} 2003, \aap, 406, L59

\bibitem[{{Pagani} {et~al.}(2005){Pagani}, {Pardo}, {Apponi}, {Bacmann}, \&
  {Cabrit}}]{Pagani05}
{Pagani}, L., {Pardo}, J.-R., {Apponi}, A.~J., {Bacmann}, A., \& {Cabrit}, S.
  2005, \aap, 429, 181

\bibitem[{{Prasad} \& {Huntress}(1980)}]{Prasad80}
{Prasad}, S.~S. \& {Huntress}, Jr., W.~T. 1980, \apjs, 43, 1

\bibitem[{{Pratap} {et~al.}(1997){Pratap}, {Dickens}, {Snell}, {Miralles},
  {Bergin}, {Irvine}, \& {Schloerb}}]{Pratap97}
{Pratap}, P., {Dickens}, J.~E., {Snell}, R.~L., {et~al.} 1997, \apj, 486, 862

\bibitem[{{Rachford} {et~al.}(2009){Rachford}, {Snow}, {Destree}, {Ross},
  {Ferlet}, {Friedman}, {Gry}, {Jenkins}, {Morton}, {Savage}, {Shull},
  {Sonnentrucker}, {Tumlinson}, {Vidal-Madjar}, {Welty}, \&
  {York}}]{Rachford09}
{Rachford}, B.~L., {Snow}, T.~P., {Destree}, J.~D., {et~al.} 2009, \apjs, 180,
  125

\bibitem[{{Ruaud} {et~al.}(2016){Ruaud}, {Wakelam}, \& {Hersant}}]{Ruaud16}
{Ruaud}, M., {Wakelam}, V., \& {Hersant}, F. 2016, \mnras, 459, 3756

\bibitem[{{Snow} \& {McCall}(2006)}]{Snow06}
{Snow}, T.~P. \& {McCall}, B.~J. 2006, \araa, 44, 367

\bibitem[{{Tafalla} {et~al.}(2004){Tafalla}, {Myers}, {Caselli}, \&
  {Walmsley}}]{Tafalla04}
{Tafalla}, M., {Myers}, P.~C., {Caselli}, P., \& {Walmsley}, C.~M. 2004, \aap,
  416, 191

\bibitem[{{van Dishoeck} {et~al.}(1993){van Dishoeck}, {Blake}, {Draine}, \&
  {Lunine}}]{vanDishoeck93}
{van Dishoeck}, E.~F., {Blake}, G.~A., {Draine}, B.~T., \& {Lunine}, J.~I.
  1993, in Protostars and Planets III, ed. E.~H. {Levy} \& J.~I. {Lunine},
  163--241

\bibitem[{{Vidal} {et~al.}(2017){Vidal}, {Loison}, {Jaziri}, {Ruaud},
  {Gratier}, \& {Wakelam}}]{Vidal17}
{Vidal}, T.~H.~G., {Loison}, J.-C., {Jaziri}, A.~Y., {et~al.} 2017, \mnras,
  469, 435

\bibitem[{{Wakelam} \& {Herbst}(2008)}]{Wakelam08}
{Wakelam}, V. \& {Herbst}, E. 2008, \apj, 680, 371

\bibitem[{{Wakelam} {et~al.}(2015{\natexlab{a}}){Wakelam}, {Loison}, {Herbst},
  {Pavone}, {Bergeat}, {B{\'e}roff}, {Chabot}, {Faure}, {Galli}, {Geppert},
  {Gerlich}, {Gratier}, {Harada}, {Hickson}, {Honvault}, {Klippenstein}, {Le
  Picard}, {Nyman}, {Ruaud}, {Schlemmer}, {Sims}, {Talbi}, {Tennyson}, \&
  {Wester}}]{Wakelam15}
{Wakelam}, V., {Loison}, J.-C., {Herbst}, E., {et~al.} 2015{\natexlab{a}},
  \apjs, 217, 20

\bibitem[{{Wakelam} {et~al.}(2015{\natexlab{b}}){Wakelam}, {Loison}, {Hickson},
  \& {Ruaud}}]{Wakelam15b}
{Wakelam}, V., {Loison}, J.-C., {Hickson}, K.~M., \& {Ruaud}, M.
  2015{\natexlab{b}}, \mnras, 453, L48

\bibitem[{{Wakelam} {et~al.}(2010){Wakelam}, {Smith}, {Herbst}, {Troe},
  {Geppert}, {Linnartz}, {{\"O}berg}, {Roueff}, {Ag{\'u}ndez}, {Pernot},
  {Cuppen}, {Loison}, \& {Talbi}}]{Wakelam10}
{Wakelam}, V., {Smith}, I.~W.~M., {Herbst}, E., {et~al.} 2010, \ssr, 156, 13

\bibitem[{{Watson}(1976)}]{Watson76}
{Watson}, W.~D. 1976, Reviews of Modern Physics, 48, 513

\bibitem[{{Whittet} {et~al.}(2001){Whittet}, {Gerakines}, {Hough}, \&
  {Shenoy}}]{Whittet01}
{Whittet}, D.~C.~B., {Gerakines}, P.~A., {Hough}, J.~H., \& {Shenoy}, S.~S.
  2001, \apj, 547, 872

\bibitem[{{Willacy} {et~al.}(1998){Willacy}, {Langer}, \&
  {Velusamy}}]{Willacy98}
{Willacy}, K., {Langer}, W.~D., \& {Velusamy}, T. 1998, \apjl, 507, L171

\end{thebibliography}

\end{document}